\documentclass[twocolumn,usenames,dvipsnames]{aastex631}

\usepackage{graphicx}
\usepackage{bm}
\usepackage{amssymb,amsmath}
\usepackage{mathrsfs} 
\usepackage{latexsym}
\usepackage{mathtools} 
\usepackage{color}
\usepackage[normalem]{ulem} 
\usepackage{tikz}
\usepackage{pgfplots}
\usepackage{hyperref}
\usepackage{natbib}
\usepackage{multirow}
\usepackage{xspace}
\usepackage[nolist,nohyperlinks]{acronym}
\usepackage{booktabs}

\usepackage[deletedmarkup=xout,commentmarkup=uwave,commandnameprefix=ifneeded]{changes}

\definechangesauthor[name=Max, color=teal]{MI}
\definechangesauthor[name=Aaron, color=purple]{AZ}
\definechangesauthor[name=Asad, color=brown]{AH}

\allowdisplaybreaks

\graphicspath{{.}}

\newcommand{\Austin}{\affiliation{Weinberg Institute, University of Texas at Austin, Austin, TX 78712, USA}}
\newcommand{\CCA}{\affiliation{Center for Computational Astrophysics, Flatiron Institute, NY}}
\newcommand{\CU}{\affiliation{
    Columbia University, Department of Astronomy, 550 West 120th Street, New York, NY, 10027, USA
}}

\newcommand{\R}[1]{\textcolor{red}{#1}}

\pgfplotsset{compat=1.18}

\begin{document}

\begin{acronym}
    \acro{GW}{gravitational wave}
    \acro{GR}{general relativity}
    \acro{CBC}{compact binary coalescence}
    \acro{BH}{black hole}
    \acro{NS}{neutron star}
    \acro{BBH}{binary black hole}
    \acro{LVK}{LIGO--Virgo--KAGRA}
    \acro{PE}{parameter estimation}
    \acro{FAR}{false-alarm rate}
    \acro{GWOSC}{the Gravitational Wave Open Science Center}
    \acro{SSB}{solar system barycenter}
    \acro{SPA}{stationary phase approximation}
    \acro{PN}{post-Newtonian}
    \acro{BNS}{binary neutron star}
    \acro{IMR}{inspiral-merger-ringdown}
    \acro{NSF}{National Science Foundation}
    \acro{GMM}{Gaussian mixture model}
    \acro{MCMC}{Markov chain Monte Carlo}
    \acro{MC}{Monte Carlo}
    \acro{BIC}{Bayesian information criterion}
	\acro{BF}{Bayes factor}
	\acro{SDDR}{Savage Dickey density ratio}
	\acro{TGMM}{truncated Gaussian mixture model}
	\acro{PPD}{population predictive distribution}
	\acro{HPDI}{highest posterior density interval}
	\acro{AGN}{active galactic nucleus}
	\acro{KDE}{kernel density estimate}
    \acro{CI}{credible interval}
    \acro{CL}{credible level}
\end{acronym}

\def\checkmark{\tikz\fill[scale=0.4](0,.35) -- (.25,0) -- (1,.7) -- (.25,.15) -- cycle;}

\newcommand{\IndependentFreeTiltMuChiOneA}{0.05^{+0.06}_{-0.05}}
\newcommand{\IndependentFreeTiltMuChiTwoA}{0.07^{+0.09}_{-0.07}}
\newcommand{\IndependentFreeTiltSigmaChiOneA}{0.07^{+0.11}_{-0.07}}
\newcommand{\IndependentFreeTiltSigmaChiTwoA}{0.11^{+0.14}_{-0.11}}
\newcommand{\IndependentFreeTiltMuChiOneB}{0.66^{+0.34}_{-0.44}}
\newcommand{\IndependentFreeTiltMuChiTwoB}{0.6^{+0.4}_{-0.43}}
\newcommand{\IndependentFreeTiltSigmaChiOneB}{0.54^{+0.46}_{-0.43}}
\newcommand{\IndependentFreeTiltSigmaChiTwoB}{0.54^{+0.45}_{-0.41}}
\newcommand{\IndependentFreeTiltMuTilt}{-2.31^{+43.97}_{-39.26} (\times 10^{-3})}
\newcommand{\IndependentFreeTiltSigmaTilt}{0.03^{+0.05}_{-0.03}}
\newcommand{\IndependentFreeTiltTiltFraction}{0.7^{+0.16}_{-0.17}}
\newcommand{\IndependentFreeTiltFraction}{0.83^{+0.11}_{-0.13}}
\newcommand{\IndependentFreeTiltFractionOneMinus}{0.17^{+0.13}_{-0.11}}
\newcommand{\IndependentFreeTiltTrueCorrelationA}{-0.04^{+0.85}_{-0.43}}
\newcommand{\IndependentFreeTiltTrueCorrelationB}{0.04^{+0.46}_{-0.39}}
\newcommand{\IndependentFreeTiltLogTenBFForZeroSpinSubpop}{3.93^{+0.48}_{-1.14}}
\newcommand{\IndependentFreeTiltLogTenBFAgainstOneSubpop}{1.53^{+0.24}_{-0.23}}

\newcommand{\MassDependentFreeTiltMuChiOneA}{0.07^{+0.06}_{-0.07}}
\newcommand{\MassDependentFreeTiltMuChiTwoA}{0.09^{+0.09}_{-0.09}}
\newcommand{\MassDependentFreeTiltSigmaChiOneA}{0.08^{+0.12}_{-0.08}}
\newcommand{\MassDependentFreeTiltSigmaChiTwoA}{0.12^{+0.14}_{-0.12}}
\newcommand{\MassDependentFreeTiltMuChiOneB}{0.75^{+0.25}_{-0.41}}
\newcommand{\MassDependentFreeTiltMuChiTwoB}{0.69^{+0.31}_{-0.44}}
\newcommand{\MassDependentFreeTiltSigmaChiOneB}{0.43^{+0.46}_{-0.43}}
\newcommand{\MassDependentFreeTiltSigmaChiTwoB}{0.49^{+0.51}_{-0.39}}
\newcommand{\MassDependentFreeTiltMuTilt}{-7.06^{+38.59}_{-39.2} (\times 10^{-3})}
\newcommand{\MassDependentFreeTiltSigmaTilt}{0.03^{+0.05}_{-0.03}}
\newcommand{\MassDependentFreeTiltTiltFraction}{0.66^{+0.16}_{-0.16}}
\newcommand{\MassDependentFreeTiltMassStart}{36^{+19}_{-21}}
\newcommand{\MassDependentFreeTiltMassEnd}{68^{+65}_{-31}}
\newcommand{\MassDependentFreeTiltFractionLowMass}{0.95^{+0.05}_{-0.08}}
\newcommand{\MassDependentFreeTiltFractionHighMass}{0.17^{+0.29}_{-0.17}}
\newcommand{\MassDependentFreeTiltTrueCorrelationA}{-0.02^{+0.84}_{-0.43}}
\newcommand{\MassDependentFreeTiltTrueCorrelationB}{0.01^{+0.51}_{-0.44}}
\newcommand{\MassDependentFreeTiltFraction}{0.95^{+0.05}_{-0.09}}
\newcommand{\MassDependentFreeTiltFractionOneMinus}{0.05^{+0.09}_{-0.05}}
\newcommand{\MassDependentFreeTiltLogTenBFForZeroSpinSubpop}{2.1^{+0.7}_{-4.5}}
\newcommand{\MassDependentFreeTiltLogTenBFAgainstOneSubpop}{4^{+6}_{-2}}
\newcommand{\MassDependentFreeTiltLogTenBFAgainstSameFraction}{1.6^{+4.0}_{-0.5}}

\title{Evidence for mass-dependent spin subpopulations in GWTC-4}

\author[0000-0003-3491-5439]{Asad Hussain}
\CCA
\author[0000-0001-8830-8672]{Maximiliano Isi}
\CU
\CCA
\author[0000-0002-7453-6372]{Aaron Zimmerman}
\Austin
\date{\today}

\begin{abstract}
While the origin of merging black-hole binaries observed in gravitational waves remain uncertain, different formation channels are expected to leave distinct imprints on their observed mass and spin distributions.
In this work, we focus on the mass dependence of the spin magnitudes $(\chi_1,\chi_2)$, allowing for multiple spin subpopulations whose relative fraction varies with mass.
Using the binaries from the fourth LIGO--Virgo--KAGRA gravitational wave transient catalog (GWTC-4), we find strong evidence for two subpopulations, with a log Bayes factor of $\log_{10} \mathcal B = \MassDependentFreeTiltLogTenBFAgainstOneSubpop$.
The data support a picture in which low-mass systems are composed almost entirely of slowly spinning black holes ($\chi \lesssim 0.2$), while high-mass systems are dominated by a broader component with moderate to rapid spins, peaking at $(\chi_1,\chi_2)\approx(0.7,0.7)$ and retaining a non-negligible tail toward asymmetric spins $(\chi_1,\chi_2)\approx(0.7,0)$.
The transition between these regimes spans a broad range from $m_1 \approx 35^{+20}_{-20} \,M_\odot$ to $m_1 \approx 70^{+90}_{-25}\,M_\odot$.
For the low-mass, slowly-spinning regime, a flexible model of spin orientations does not allow us to exclude identically vanishing spins, $\chi_1 = \chi_2 = 0$.
Meanwhile, the high-mass, rapidly-spinning population does not neatly display the spin-magnitude asymmetry expected from hierarchical mergers, although the data do not exclude that.
This kind of localized structure is generally difficult to access with standard Monte-Carlo likelihood estimators, which we replace with a regularized representation in terms of truncated Gaussian mixtures.
Taken together, our results provide a new target for formation models, which should reproduce not only the mass and effective spin distributions, but also the mass-dependent structure of the individual component-spin magnitudes.
\end{abstract}

\section{Introduction}
Over the past decade, \ac{GW} detections by the Advanced LIGO~\citep{LIGOScientific:2014pky}, Virgo~\citep{VIRGO:2014yos} and KAGRA \citep{PhysRevD.88.043007} detectors have provided a powerful probe of astrophysics \citep{LIGOScientific:2025pvj}.
The origin of the binaries that produce these signals, composed of \acp{BH} and \acp{NS}, remains an open question.
Answering it addresses the complex and still uncertain physics underlying the lives and deaths of massive stars, especially in binary systems \citep[e.g.,][]{Mandel:2018hfr}.
With more than 200 signals detected~\citep{LIGOScientific:2025slb}, primarily from \acp{BBH}, population-level inference has revealed multiple features in the mass distribution of \acp{BH}~\citep{LIGOScientific:2020kqk,Tiwari:2020otp,Edelman:2021zkw}, indications that multiple channels produce merging binaries~\citep{LIGOScientific:2020kqk}, possible signatures of pair-instability processes~\citep{Tong:2025wpz, Ray:2025xti}, and correlations among binary parameters that promise further insights~\citep{LIGOScientific:2025pvj}.

Remarkably, the spins of compact objects are directly measured from the \ac{GW} signals.
The orientation of these spins relative to the binary orbit, encoded in their tilt angles $\theta_i$ can reveal the imprint of interactions with other bodies~\citep{2017MNRAS.471.2801S, 2018PhRvD..98h4036G}.
The (dimensionless) magnitudes of the spins $\chi_i$ hold information about angular momentum and its transport in the cores of massive stars~\citep{Fuller:2019sxi, Ma:2019cpr}, as well as angular momentum transfer from a stellar companion~\citep{Qin:2018vaa, Ma:2023nrf}.
It can also reveal a population of second-generation binaries \citep[e.g.,][]{2021NatAs...5..749G}, one or both of whose members was formed in a previous \ac{BBH} merger and so would have $\chi\sim 0.7$, as determined by the orbital angular momentum of the binary \citep{Pretorius:2005gq, Scheel:2008rj}.
Hints of multiple populations in the \ac{BBH} spin distribution are already present in the GWTC-3~\citep{ligo_scientific_collaboration_and_virgo_2021_5636816} and GWTC-4 \citep{LIGOScientific:2025pvj} catalogs, indicating subpopulations with either high or low \ac{BH} spins ~\citep{Galaudage2021, Tong2022, Hussain2025, Bartos:2026xlt, Banagiri:2025dmy, Ray:2026uur}, and previous population studies have also have found signs of possible correlations between black-hole masses and spins in merging \acp{BBH}~\citep{Franciolini:2022iaa, Adamcewicz:2022hce, LIGOScientific:2025pvj, Pierra:2024fbl, Mould:2022ccw}.
Reduced observables such as $\chi_\mathrm{eff}$ can, however, mix together physically distinct spin configurations, obfuscating astrophysical scenarios and motivating analyses in component-spin space \citep{Payne:2024ywe}.

Inferring the distribution of \ac{BH} spins can be challenging technically, especially for models which support multiple populations and where the distributions are allowed to be narrow, as when allowing for a concentration at negligible spins.
Standard \ac{GW} population methods use samples derived from Bayesian \ac{PE} produced with a fiducial prior to estimate the population-level likelihood via \ac{MC} averaging.
These approaches are often limited by the variance of such estimators~\citep{Talbot:2023pex, Heinzel:2025ogf, Hussain:2025llf}, which becomes especially severe for narrow populations.
Addressing this limitation with specialized approaches is key to further improving our understanding of the population of \ac{GW} sources, particularly as our catalogs grow \citep[e.g.][]{Essick:2022ojx,Talbot:2023pex, Heinzel:2025ogf}.
As we showed in \citet{Hussain2025,Hussain:2025llf}, these issues can be mitigated by using (truncated) Gaussian mixture models to regularize the likelihood.

In~\citet{Hussain2025}, we studied the distribution of \ac{BH} spin magnitudes in GWTC-3 by allowing for correlations in the space of component spin magnitudes, with up to two independent subpopulations in that space.
The result was tentative evidence for two distinct spin distributions:
a dominant distribution displayed primary spins $\chi_1\sim 0.2$, smaller secondary spins, and some potential anti-correlation between the two;
a subdominant distribution, containing ${\sim} 15\%$ of the \acp{BBH}, preferred high spins $\chi_i \sim 0.8$ for both \acp{BH}, with indications of preferentially containing more massive binaries.
Some of these results were driven by GW190517, a binary with large component spins well-measured away from zero~\citep{LIGOScientific:2020ibl,LIGOScientific:2021usb}.
In this work we update \citet{Hussain2025} using GWTC-4 and introduce an explicit mass dependence to the mixing fraction between the two subpopulations.

Altogether, our results suggest distinct formation channels are in operation across \acp{BBH} masses.
In Sec.~\ref{sec:Methods} we discuss our population models and dataset.
We present our inferences in Sec.~\ref{sec:Results}, and discuss their astrophysical implications in Sec.~\ref{sec:Implications}.
We conclude in Section~\ref{sec:Conclusions}. 
In App.~\ref{sec:MethodsAppx} we provide greater detail into some aspects of our modeling and methods, and in App.~\ref{sec:ResultsAppx} we present additional population inference results.

\section{Methods and models}
\label{sec:Methods}

\nolinenumbers
\begin{figure}[t]
    \centering
    \begin{tikzpicture}%
    \begin{axis}[%
        width=8cm, height=6cm,
        xmin=0, xmax=10,
        ymin=0, ymax=1.2,
        axis lines=left,
        xlabel={$m_1$},
        ylabel={$\eta(m_1)$},
        xtick={3, 7},
        xticklabels={$m_s$, $m_e$},
        ytick={0.8, 0.3},
        yticklabels={$\eta_s$, $\eta_e$},
        tick align=outside,
        clip=false,
    ]%
    \addplot[thick, blue, domain=0:3] {0.8};
    \addplot[thick, blue, domain=3:7] {0.8 + (0.3 - 0.8)/(7 - 3) * (x - 3)};
    \addplot[thick, blue, domain=7:10] {0.3};
    \addplot[dashed, gray] coordinates {(3, 0) (3, 0.8)};
    \addplot[dashed, gray] coordinates {(0, 0.8) (3, 0.8)};
    \addplot[dashed, gray] coordinates {(7, 0) (7, 0.3)};
    \addplot[dashed, gray] coordinates {(0, 0.3) (7, 0.3)};
    \end{axis}%
    \end{tikzpicture}%
    \caption{Functional form of the mass-dependent fraction $\eta(m_1)$. The fraction remains constant at $\eta_s$ for $m_1 < m_s$, transitions linearly to $\eta_e$ between $m_s$ and $m_e$, and remains constant at $\eta_e$ for $m_1 > m_e$.}
    \label{fig:eta_cartoon}
\end{figure}
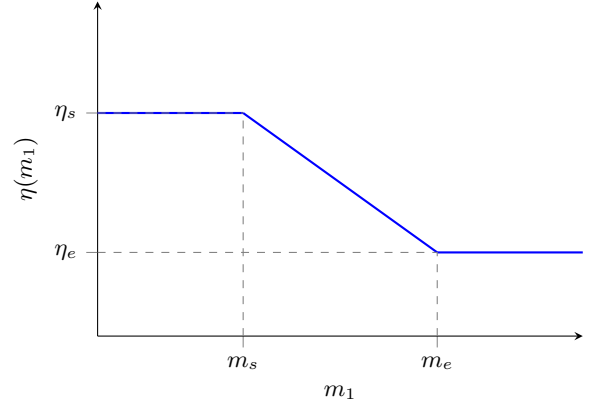

We use hierarchical Bayesian inference to infer the population properties of \acp{BBH}~\citep{Mandel:2018mve,Thrane:2018qnx,Vitale:2020aaz,LIGOScientific:2025pvj}. 
The goal is to compute posteriors over the hyper-parameters $\boldsymbol{\Lambda}$ of our chosen population model.
We take as our baseline the fiducial strongly modeled \ac{BBH} population model applied in~\cite{LIGOScientific:2025pvj}: the primary mass $m_1$ drawn from a broken power law plus two normal distributions (``peaks''), the mass ratio $q = m_2/m_1$ drawn from a power law conditioned on $m_1$, and a merger rate that evolves as a power law in $1+z$, with $z$ the redshift.

We adopt a flexible, two-subpopulation model for the \ac{BBH} spin magnitudes, $\bm{\chi}=(\chi_1, \chi_2)$, similar to \citealt{Hussain2025}, drawing them from a mixture of two correlated and truncated 2D normal distributions (indexed by $a$ and $b$) now, unlike \citet{Hussain2025}, with a mixing fraction $\eta(m_1)$ that can depend on the primary mass $m_1$. 
The model is
\begin{align}
	p(\bm{\chi}) &= \eta(m_1) {\mathcal N}_{[\boldsymbol{0},\boldsymbol{1}]}
	\left(\bm{\chi} \mid \bm{\mu}^a, \boldsymbol{\Sigma}^a\right) \nonumber\\
	&+\left[1-\eta(m_1)\right] {\mathcal N}_{[\boldsymbol{0},\boldsymbol{1}]}
	\left(\bm{\chi} \mid \bm{\mu}^b, \boldsymbol{\Sigma}^b\right),
	\label{eq:twocomponentmodel}
\end{align}
where the $[\boldsymbol{0},\boldsymbol{1}]$ subscript indicates truncation of our domain to the $[0,1]\times[0,1]$ unit square, while both $\bm{\Sigma}^a$ and $\bm{\Sigma}^b$ independently have the general form
\begin{equation}
	\bm{\Sigma}  = \begin{pmatrix}
		\sigma_1^2 & \rho \sigma_1 \sigma_2 \\
		\rho \sigma_1 \sigma_2 &  \sigma_2^2
	\end{pmatrix} .
\end{equation}
We adopt broad, uniform hyperpriors on the spin-magnitude locations, widths, and correlations, with the prior on $\eta_s$ restricted to $[0.5,1]$. Full details are given in Appendix~\ref{sec:MethodsAppx} and Table~\ref{tab:hyperpriors}.

In addition to the spin magnitudes, the cosine tilt of each \ac{BH} spin relative to the orbital angular momentum plays a key role in determining how spins impact the GW signal.
Our fiducial model is that of~\cite{LIGOScientific:2025pvj}, where both spin tilts are drawn from either an isotropic component (uniform in cosine tilt) or a truncated normal distribution whose peak is free to move (the \textsc{free} model).
We also infer the \ac{BH} spin distribution using a similar model~\citep{KAGRA:2021duu} where the peak of the truncated normal is fixed to $\cos \theta_i = 1$ (the \textsc{fixed} model), since this was the model adopted in~\cite{Hussain2025}, and discuss our inferences using this model in Appendix~\ref{sec:MethodsAppx}.

Motivated by hints that massive black holes lie within the highly spinning subpopulation~\citep{Hussain2025}, we allow the mixing parameter to vary from a constant fraction $\eta_s$ at low masses to a constant value $\eta_e$ for high masses, linearly interpolating between them in a transitional regime between $m_s$ and $m_e$; see Figure~\ref{fig:eta_cartoon}.
\added[id=AZ]{Here we fix $\eta_s \geq 0.5$ so that $a$ is the population that dominates at low masses.}
The functional form of $\eta(m_1)$, and further information on our population model and hyper-priors is given in the Appendix.
We also carry out inference with a constant mixing fraction $\eta$, although this is disfavored relative to the mass-dependent model, for comparison with~\cite{Hussain2025}.

We use the \added[id=AZ]{153} \ac{BBH} candidate signals which have \ac{FAR} $< 1\, \mathrm{yr}^{-1}$ in GWTC-4~\citep{LIGOScientific:2025slb, LIGOVirgoKAGRA:2025gwtc4pe,LIGOScientific:2025snk} in our inferences.
For each event, we start from the \textsc{Mixed} posterior samples from GWTC-4 \citep{LIGOScientific:2021usb,ligo_scientific_collaboration_and_virgo_2021_5636816,LIGOVirgoKAGRA:2025gwtc4pe}, 
derived by combining \ac{PE} results across multiple waveform models~\citep{LIGOScientific:2025yae}.
We account for selection effects using the semi-analytic sensitivity estimates and injection-recovery campaign used by \citet{LIGOScientific:2025pvj, LIGOVirgoKAGRA:2025gwtc4sens}
and described in~\cite{Essick:2025zed,LIGOScientific:2025yae, Essick:2023toz}

Evaluating the population likelihood $\mathcal L(\boldsymbol{\Lambda}| \{d\})$ requires integrating the population model over the likelihood for each event.
Similarly, accounting for selection effects using the detection fraction $\xi(\boldsymbol{\Lambda})$ requires integrating the population model over the detection probability for binaries with given parameters.
These integrals are typically carried out by summing over samples from reference distributions using \ac{MC} methods, and are limited by the variance of the \ac{MC} integrals~\citep{Thrane:2018qnx, Mandel:2018mve}, particularly for narrow population distributions which may otherwise be consistent with the data~\citep{Talbot:2023pex, Heinzel:2025ogf}.
To address this we use the methods described in~\citet{Hussain2025,Hussain:2025llf}, fitting \acp{TGMM} to the underlying (weighted) event posteriors or recovered injections.
We carry out the needed integrals efficiently using a combination of semi-analytic integration (in the spin sector of parameter space) and \ac{MC} integration (in the mass and redshift sector) using the \acp{TGMM}, controlling the variance even for $\boldsymbol{\Lambda}$ values where the spin subpopulations are narrow or concentrated in corners of parameter space.
We briefly review the method and updates made for this work in Appendix~\ref{sec:MethodsAppx}.

We carry out the \ac{TGMM} fitting using the \textsc{truncatedgaussianmixtures} package~\citep{truncatedgaussianmixtures} and population inference using the \textsc{gravpop} package~\citep{gravpop}.
We sample our hyper-posteriors using the no-U-turn sampler \citep{2011arXiv1111.4246H} Hamiltonian \ac{MC} algorithm \citep{2011hmcm.book..113N,2017arXiv170102434B} implemented in \textsc{numpyro}~\citep{phan2019composable,bingham2019pyro}.
\section{Results}
\label{sec:Results}

\begin{figure*}[tb]
    \centering
    \includegraphics[width=1.20\columnwidth]{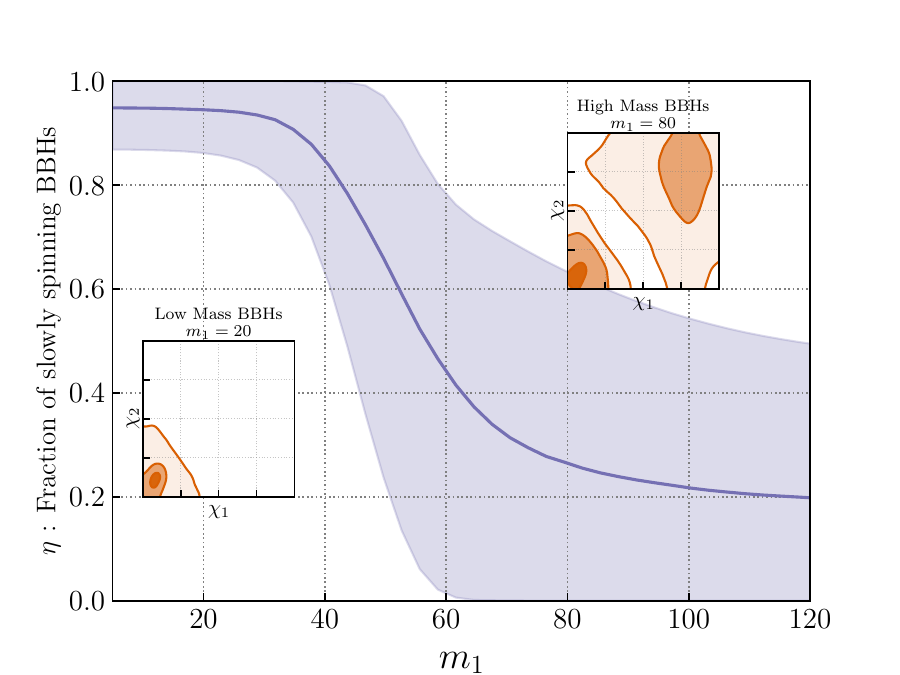}
    \includegraphics[width=0.82\columnwidth]{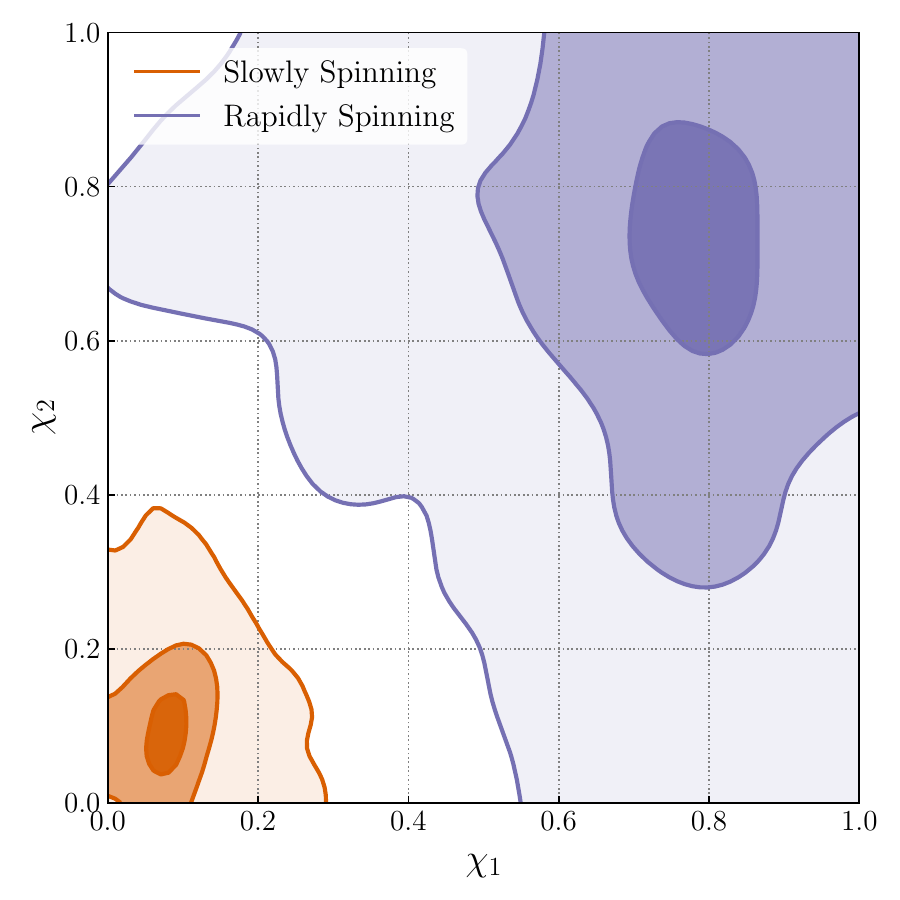}
    \caption{
    {\it Left:} \ac{PPD} of the mixing fraction $\eta(m_1)$ showing a transition at higher masses. The insets show the inferred spin magnitude distribution at $m_1 = 20\, M_\odot$ and at $80\,M_\odot$, below and above the transition.
    {\it Right:} The inferred spin magnitude distribution for each of the two components, individually normalized. The slowly spinning subpopulation is the one that dominates at low masses. The rapidly spinning subpopulation peaks in the region where both \acp{BH} have large spins, but this distribution is diffuse and has support in regions where either the primary or secondary have small spins.
    }
    \label{fig:mass_dependent_fraction_ppd}
\end{figure*}

Incorporating the additional 84 above-threshold \acp{BBH} in the GWTC-4 dataset, we find strong evidence not only for the presence of two spin-magnitude subpopulations but also for a mass-dependent mixing fraction between them. 
Figure~\ref{fig:mass_dependent_fraction_ppd} shows our inferred distribution for the mixing fraction  $\eta(m_1)$ between the two spin-magnitude subpopulations as a function of primary mass $m_1$.
We find evidence for a transition in the mixture fraction beginning at $m_s = \MassDependentFreeTiltMassStart \, M_\odot$ and ending at $m_e = \MassDependentFreeTiltMassEnd \, M_\odot$, transitioning from $\eta_s = \MassDependentFreeTiltFractionLowMass$ to $\eta_e = \MassDependentFreeTiltFractionHighMass$.
From the Savage-Dickey density ratio at $\eta_s = \eta_e = 1$, we compute the log \ac{BF} for two versus one subpopulation to be 
$\log_{10} \mathcal B = \MassDependentFreeTiltLogTenBFAgainstOneSubpop$, relying on a 2D \ac{KDE} suitable for square domains.

The component that dominates at lower masses is tightly concentrated towards small spin magnitudes, with $\boldsymbol{\mu}^a = (\MassDependentFreeTiltMuChiOneA,\MassDependentFreeTiltMuChiTwoA)$ and $\boldsymbol{\sigma}^a = (\MassDependentFreeTiltSigmaChiOneA, \MassDependentFreeTiltSigmaChiTwoA)$.
This is illustrated in the right panel of Figure~\ref{fig:mass_dependent_fraction_ppd}, which shows the two inferred spin distributions individually normalized.
At higher masses, $\eta(m_1)$ is distributed more broadly, so that \acp{BBH} arise from both the highly spinning and slowly spinning subpopulations, although the former most likely dominates.
As we found previously with GWTC-3 data, this subpopulation prefers large spin magnitudes, although it is more diffuse than the slowly spinning spin distribution, with $\boldsymbol{\mu^b} = (\MassDependentFreeTiltMuChiOneB,\MassDependentFreeTiltMuChiTwoB)$ and $\boldsymbol{\sigma}^b = (\MassDependentFreeTiltSigmaChiOneB, \MassDependentFreeTiltSigmaChiTwoB)$.

\begin{figure*}[tb]
     \centering
     \includegraphics[width=0.85\columnwidth]{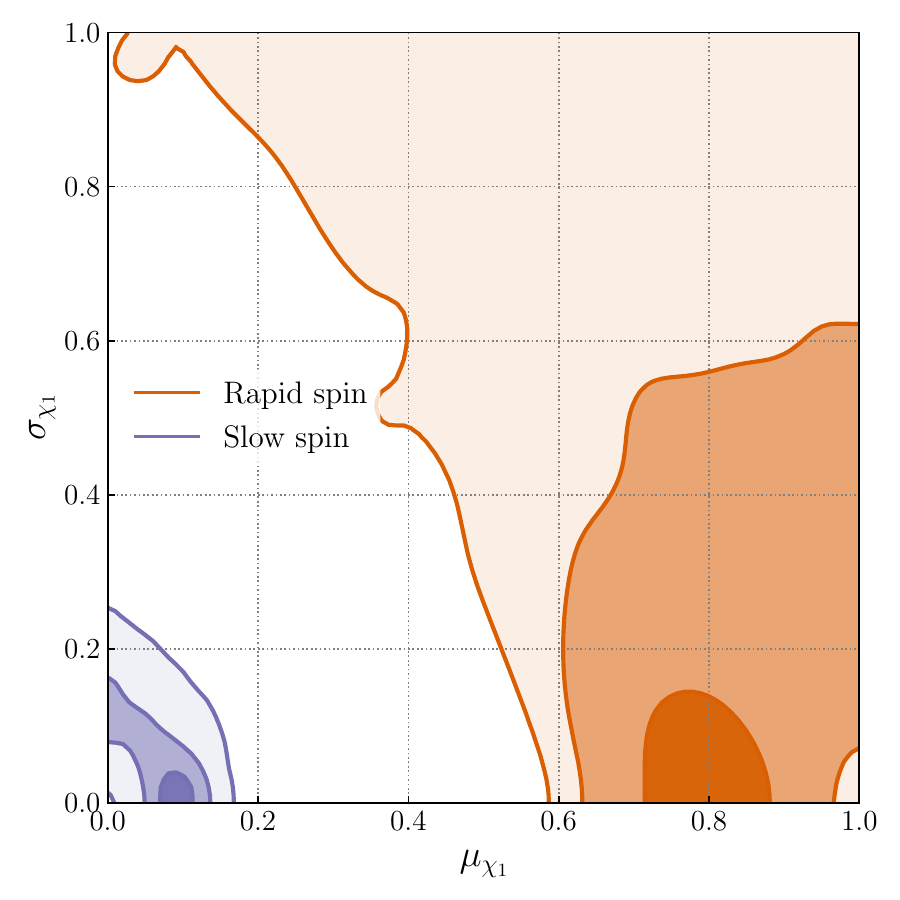} 
     \qquad
     \includegraphics[width=0.85\columnwidth]{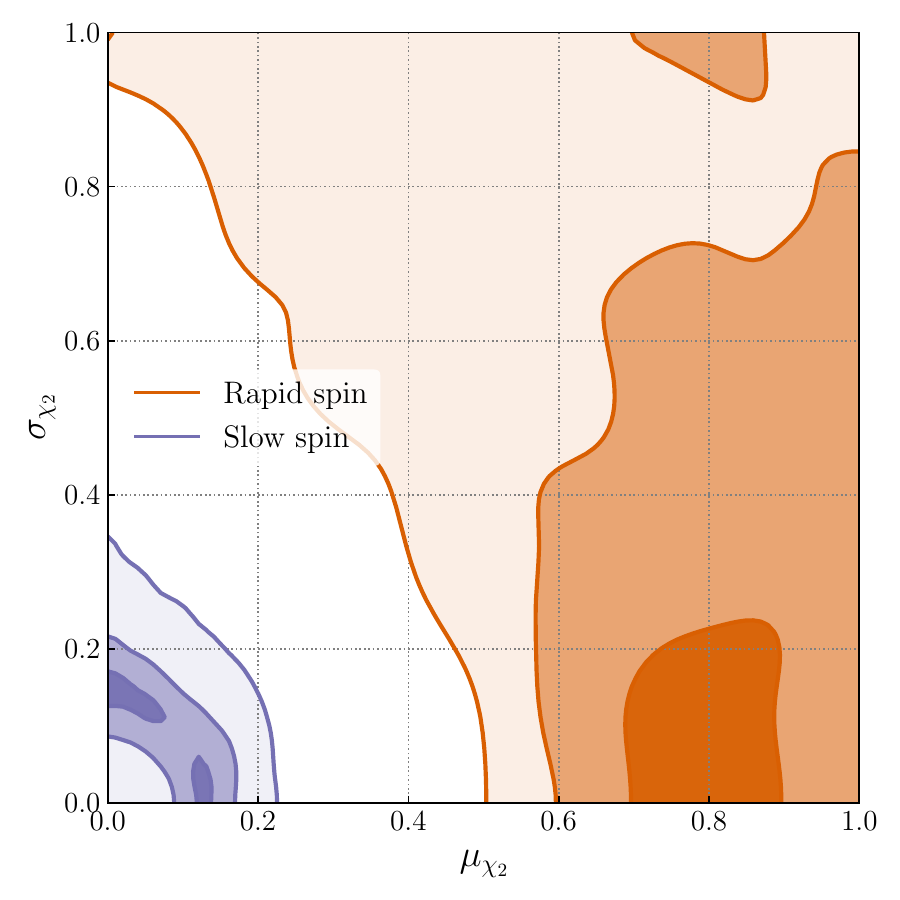}
     \caption{Posteriors over the hyperparameters of the spin magnitude population model, split by the location and scale hyperparameters of the primary \ac{BH} (\textit{left}) and the secondary \ac{BH} (\textit{right}). 
     The slowly spinning subpopulation has $\boldsymbol{\mu}^a = (\MassDependentFreeTiltMuChiOneA,\MassDependentFreeTiltMuChiTwoA)$, and $\boldsymbol{\sigma}^a = (\MassDependentFreeTiltSigmaChiOneA, \MassDependentFreeTiltSigmaChiTwoA)$.
     The rapidly spinning subpopulation has $\boldsymbol{\mu}^b = (\MassDependentFreeTiltMuChiOneB,\MassDependentFreeTiltMuChiTwoB)$, with 
    $\boldsymbol{\sigma}^b = (\MassDependentFreeTiltSigmaChiOneB, \MassDependentFreeTiltSigmaChiTwoB)$, preferring that both \acp{BH} are spinning rapidly.
     }
     \label{fig:spin_hyper_posterior}
 \end{figure*}

The features of the two spin-magnitude distributions are in broad agreement with our previous results.
Figure~\ref{fig:spin_hyper_posterior} shows the inferred posteriors over the hyperparameters controlling the spin distributions.
One difference relative to \cite{Hussain2025} is that, for the slowly spinning population, we find some support for all \acp{BH} in this population having vanishing spins, i.e., $\boldsymbol{\mu^a} = (0,0)$ and $\boldsymbol{\sigma}^a = (0,0)$, which was disfavored in our GWTC-3 analysis.
The log \ac{BF} for this negligible-spin hypothesis is $\log_{10}\mathcal B = \MassDependentFreeTiltLogTenBFForZeroSpinSubpop$, encompassing positive and negative log \ac{BF} within our estimated uncertainties.
This difference compared to our past work is due in part to our more flexible spin-tilt model: we now infer the peak of the truncated normal to be near $\cos\theta_i = 0$, allowing for spins near the orbital plane.
If we repeat our analysis while fixing the peak to be at $\cos \theta_i = 1$, the slowly spinning distribution moves to higher spin magnitudes, as discussed in Appendix~\ref{sec:ResultsAppx}.
This is likely indicative of further substructure within the low spin population.

The rapidly-spinning subpopulation, meanwhile, includes greater support for the case in which one of the two \acp{BH} has small spin while the other has $\chi_i \sim 0.7$, a characteristic feature of a hierarchical merger between a second- and first-generation \ac{BH}.
This feature is independent of our choice of tilt model.
Overall though the population hyperparameters still favor populations where both \acp{BH} have relatively large spins.

Our model for spin magnitudes allows for correlations between the primary and secondary spin magnitudes within each population. 
We compute the Pearson correlation coefficient between the spins within each subpopulation and find weak evidence that, in the slowly spinning population, the \ac{BH} spins are anti-correlated, as shown in Appendix~\ref{sec:ResultsAppx}.
Meanwhile the correlation coefficient for the rapidly spinning distribution is unchanged from the prior.

Using only the events from GWTC-3, in \citet{Hussain2025} we found that the evidence for the rapidly spinning subpopulation hinges on the \ac{BBH} with the most extreme and tightly measured spin-magnitudes, GW190517.
With the additional events from GWTC-4, our overall conclusions are not changed by a leave-one-out analysis where we remove the highly-spinning and massive \ac{BBH} GW231123~\citep{LIGOScientific:2025rsn}, nor by leaving out both GW231123 and GW190517.
Removing this event slightly broadens the rapidly spinning distribution, with marginally more support in the tail where $\chi_1$ is large and $\chi_2$ is small.
More information about our inferences when excluding these events are given in Appendix~\ref{sec:ResultsAppx}.

Our model disfavors the case of a constant $\eta$ across mass, with a log \ac{BF} against the hypothesis $\eta_s = \eta_e$ of $\log_{10} \mathcal{B} = \MassDependentFreeTiltLogTenBFAgainstSameFraction$.
Nevertheless, to facilitate comparison to the GWTC-3 analysis of~\cite{Hussain2025}, we additionally carry out inference with constant $\eta$, using the \textsc{free} spin-tilt model. 
In that case, we find $\eta = \IndependentFreeTiltFraction$ with no appreciable support for a single spin-magnitude population, i.e., $\eta = 1$ is excluded.
The inferred properties of the dominant and subdominant spin distributions in this constant-mixing-fraction model are in broad agreement with those from our fiducial mass-dependent model.
We find for the slowly spinning \acp{BH}
$\boldsymbol{\mu}^a = (\IndependentFreeTiltMuChiOneA,\IndependentFreeTiltMuChiTwoA)$, favoring narrow 
distributions of widths $\boldsymbol{\sigma}^a = (\IndependentFreeTiltSigmaChiOneA, \IndependentFreeTiltSigmaChiTwoA)$.
For the rapidly spinning distribution, we find 
$\boldsymbol{\mu}^b_1 = (\IndependentFreeTiltMuChiOneB,\IndependentFreeTiltMuChiTwoB)$, and a wide distribution with $\boldsymbol{\sigma}^b = (\IndependentFreeTiltSigmaChiOneB, \IndependentFreeTiltSigmaChiTwoB)$ consistent with our fiducial analysis.
Further results from this analysis are given in Appendix~\ref{sec:ResultsAppx}.

\section{Astrophysical implications}
\label{sec:Implications}

The mass-dependent spin mixing we infer provides a new empirical target for models of \ac{BBH} formation. 
Our results point toward a mixture of channels whose relative contribution shifts across the \ac{BBH} mass spectrum, but the current data do not uniquely identify which channels are responsible. 
Successful formation models should reproduce not only the marginal mass distribution and the distribution of effective inspiral spin, but the full conditional distribution $p(\chi_1, \chi_2 \mid m_1)$, including the transition we observe near $m_1 \sim 35\,M_\odot$.
Here we discuss how our results compare to astrophysical expectations and to recent
independent population analyses of GWTC-4.

At low primary masses, the \ac{BBH} population is concentrated near small spin magnitudes, $\chi_i \lesssim 0.1$, with posterior support extending toward zero spin. 
This is broadly consistent with isolated binary-star evolution, in which efficient angular-momentum transport from the stellar core to the envelope leaves the resulting black holes slowly rotating \citep{Fuller:2019ckz,Belczynski:2017gds}. 
Several independent analyses of GWTC-4 find a low-spin or small-$\chi_\mathrm{eff}$ subpopulation at lower masses
\citep[e.g.,][]{Banagiri:2025dmy, Sridhar:2025kvi, Ray:2026uur}, all broadly consistent with an isolated-binary origin. 
The degree to which this population is
truly negligible-spin, versus merely slowly spinning, remains uncertain.
\cite{Adamcewicz:2025phm} find no evidence for a non-spinning subpopulation in GWTC-4, while we find that the data is unable to exclude this hypothesis; this difference may reflect the mass-dependent mixing our model allows, as well as differences in the choice of spin-tilt model and in how \ac{MC} variance is controlled~\citep[see e.g.][]{Hussain:2025llf}.

Above the transition, the rapidly spinning component contributes an increasing fraction of \acp{BBH}. 
Its distribution is centered near $\chi_i \sim 0.7$ on both \acp{BH}, the characteristic remnant spin from a previous \ac{BBH} merger \citep[e.g.,][]{Fishbach:2017dwv, Gerosa:2017kvu}, although the distribution is broad and includes support for configurations where the primary spins rapidly while the secondary spins more slowly. 
The former configuration ($\chi_1 \sim \chi_2
\sim 0.7$) would be suggestive of second-generation plus second-generation (2G+2G) hierarchical pairings; the latter ($\chi_1 \sim 0.7, \chi_2 \sim 0$) resembles a 2G+1G event, a hallmark signature of hierarchical growth \citep{2021NatAs...5..749G}. 
Environments capable of retaining post-kick merger remnants---globular clusters, nuclear star clusters, or \ac{AGN} disks---are natural candidates for hosting these \acp{BBH}~\citep{Yang2019, Tagawa2021, 2021NatAs...5..749G}.
On the other hand, chemically homogeneous evolution is also consistent with the preference for both spins in the binary having high spin and high mass~\citep{Mandel:2015qlu, deMink:2016vkw}, although it does not naturally account for the asymmetric-spin tail that our inference allows.
Multiple analyses of GWTC-4 independently find evidence for a high-spin, high-mass subpopulation and variously associate it with hierarchical mergers in star clusters \citep{Antonini:2024het, Banagiri:2025dmy, Ray:2026uur}, \ac{AGN} accretion \citep{Bartos:2026xlt, Li:2025iux}, or a combination thereof; related structure in reduced spin observables has also been discussed in \cite{Vijaykumar:2026zjy}, and data-driven analyses similarly support multiple spin subpopulations \citep{Guttman:2025jkv}; our results are consistent with all of these interpretations and do not presently discriminate between them.

The transition mass $m_s = \MassDependentFreeTiltMassStart\,M_\odot$ sits in the same broad mass range where several independent studies have identified a change in character of the \ac{BBH}
population. 
\cite{Tong:2025wpz} find evidence for a pair-instability gap in the \emph{secondary} mass distribution of GWTC-4 with a lower boundary near $44\,M_\odot$, and note that binaries with primary masses in this range tend to spin more rapidly---a direct parallel to our findings. 
The absence of this gap in the primary mass distribution is naturally explained if a hierarchically formed subpopulation begins to contribute above the pair-instability scale, filling the depleted region and carrying the high-spin signature we observe \citep{Antonini:2024het, Tong:2025wpz};
\citet{Ray:2025xti} offer a complementary view, arguing that the apparent secondary-mass gap is better described as a smooth falloff, consistent with a gradual onset of hierarchical contributions rather than a sharp threshold.
This picture may also connect naturally to the second-generation peak near $70\,M_\odot$ predicted for cluster hierarchies plus a pair-instability cutoff in \citet{Ginat:2026awh}.
Similarly, \citet{Mould:2026sww} show that an inferred gradual secondary-mass falloff can remain consistent with an underlying sharp pair-instability cutoff when sufficiently flexible population models are used.
\cite{Banagiri:2025dmy} and \cite{Sridhar:2025kvi} independently find that the $35\,M_\odot$ feature in the primary mass distribution is associated with a subpopulation having isotropic spins and near-equal mass ratios---properties consistent with dynamical assembly in globular clusters---while the higher-mass subpopulation they identify has a distinctly different spin character, broadly aligned with our rapidly spinning component.

Our results underscore the importance of working in component-spin space rather than solely with
$\chi_\mathrm{eff}$. 
Different configurations of $(\chi_1, \chi_2)$---both modest, one fast and one slow, or both large but misaligned---can project onto overlapping $\chi_\mathrm{eff}$ distributions after marginalizing over tilts and mass ratios \citep{Banagiri:2025dxo, Payne:2024ywe}.
The mass-dependent $p(\chi_1, \chi_2 \mid m_1)$ we infer provides a sharp, more direct diagnostic for distinguishing formation channels, and is complementary to the $\chi_\mathrm{eff}$-based analyses of \cite{Banagiri:2025dxo} and \cite{Adamcewicz:2023szp}, who respectively find non-Gaussian structure in the $\chi_\mathrm{eff}$ distribution and that the primary \ac{BH} tends to drive the spin signal. 

Taken together, these independent lines of evidence from GWTC-4 consistently point to the operation of at least two distinct formation channels across the \ac{BBH} mass spectrum. 
The challenge for population synthesis is to reproduce the full conditional structure: a narrow low-spin component at low masses, a rapidly spinning component whose contribution grows above $m_1 \sim 35\,M_\odot$, and either saturates or coexists with the other at high mass.
The increase in the number of detected \ac{BBH} mergers promises to substantially sharpen these constraints as the size of our catalog increases.

\section{Conclusions}
\label{sec:Conclusions}

We have investigated the distribution of spin magnitudes using the 153 \acp{BBH} candidates in GWTC-4 with \ac{FAR} $<1 \,\mathrm{yr}^{-1}$.
We find strong support for at least two distinct populations of spinning \acp{BBH}, a slowly spinning subpopulation that dominates at low masses, and a rapidly spinning subpopulation that makes up a substantial fraction $\eta_e = \MassDependentFreeTiltFractionHighMass$ at high masses.
Our results point to multiple formation channels for LIGO--Virgo--KAGRA \acp{BBH} and, in particular, prefer that both black holes have large spins $\chi_i \sim 0.7$ in a significant number of high-mass \acp{BBH}. 
Such systems, like GW231123, and the asymmetric, high-spin events GW241011 and GW241110~\citep{LIGOScientific:2025brd}, pose a challenge for current models of binary formation, and do not neatly fit expectations for hierarchical mergers.
Meanwhile, the population of slowly spinning black holes can potentially be explained if a large fraction of \acp{BH}, in isolated or dynamical environments, form with small natal spins.
Relative to our GWTC-3 analysis \citep{Hussain2025}, we find restored support for a subpopulation of \acp{BBH} with vanishing spins ($\chi_i \sim 0$).
This could be due to efficient angular momentum transfer from the core to envelope during stellar evolution~\cite{Fuller:2019sxi, Ma:2019cpr}.

There are a number of natural extensions to our work that could provide further insights into \ac{BBH} formation channels and the underlying astrophysical processes.
Our population model factorizes into independent mass, redshift, spin-magnitude, and spin-tilt sectors.
However, the known features across the \ac{BH} mass spectrum, such as  the overabundance of \acp{BH} at ${\sim} 10 \, M_\odot$, and the feature at ${\sim} 35 \, M_\odot$ \citep{LIGOScientific:2025pvj}, hint at distinct \ac{BBH} formation processes in action at these masses. 
There are also additional hints of dynamical formation~\citep{LIGOScientific:2020kqk,KAGRA:2021duu,LIGOScientific:2025pvj} and systems that may be the merger of second and first generation \acp{BH}~\citep{LIGOScientific:2025brd}.
Each formation scenario is accompanied by particular expected spin magnitude and tilt distributions, and a natural next step is to couple distinct mass and tilt models to our flexible spin magnitude distributions. 
Moreover, the sensitivity of our inferred slowly-spinning population to the chosen form of our spin-tilt model indicates that further structure may be present in the \ac{BH} spin distribution, requiring additional components in our model.
Our novel \ac{TGMM} methods can help mitigate the impact of variance in likelihood estimation for such complex models, and help to reveal the imprint of specific astrophysical scenarios on \ac{BBH} populations.

Finally, our analysis was carried out using \ac{GW} detections through only the first part of the fourth LIGO-Virgo-KAGRA observing campaign. 
An additional 186 \ac{GW} candidate detections have been announced as open alerts over the remainder of the campaign~\citep{OPA}.
This large addition of binary mergers to our data set promises to sharpen our ability to infer multiple subpopulations and ultimately tease out features to help reveal the uncertain origin of these explosive events.
\\

\noindent 
We thank Jack Heinzel for valuable comments on the implementation of our methods in \textsc{gravpop}. We additionally thank Aditya Vijaykumar, Soumendra Kishore Roy, Barry McKernan and Saavik Ford for fruitful discussions about the content of this work.
This work carries preprint numbers LIGO-P2600256 and UT-WI-17-2026.
AZ was supported by NSF Grant PHY-2308833.
The Flatiron Institute is a division of the Simons Foundation.
This material is based upon work supported by NSF's LIGO Laboratory which is a major facility fully funded by the National Science Foundation.
The authors are grateful for computational resources provided by the LIGO Laboratory and supported by NSF Grants PHY-0757058 and PHY-0823459.
This research has made use of data or software obtained from the Gravitational Wave Open Science Center (gwosc.org), a service of the LIGO Scientific Collaboration, the Virgo Collaboration, and KAGRA. 
This material is based upon work supported by NSF's LIGO Laboratory which is a major facility fully funded by the National Science Foundation, as well as the Science and Technology Facilities Council (STFC) of the United Kingdom, the Max-Planck-Society (MPS), and the State of Niedersachsen/Germany for support of the construction of Advanced LIGO and construction and operation of the GEO600 detector. 
Additional support for Advanced LIGO was provided by the Australian Research Council. 
Virgo is funded, through the European Gravitational Observatory (EGO), by the French Centre National de Recherche Scientifique (CNRS), the Italian Istituto Nazionale di Fisica Nucleare (INFN) and the Dutch Nikhef, with contributions by institutions from Belgium, Germany, Greece, Hungary, Ireland, Japan, Monaco, Poland, Portugal, Spain. 
KAGRA is supported by Ministry of Education, Culture, Sports, Science and Technology (MEXT), Japan Society for the Promotion of Science (JSPS) in Japan; National Research Foundation (NRF) and Ministry of Science and ICT (MSIT) in Korea; Academia Sinica (AS) and National Science and Technology Council (NSTC) in Taiwan.

 \textit{Software: } This work was implemented using \texttt{python }\citep{Python3} and  \texttt{Julia} \citep{Julia-2017}. 
 We made use of the following python packages: \texttt{numpy }\citep{numpy},  \texttt{scipy }\citep{scipy},  \texttt{pandas }\citep{pandas},  \texttt{matplotlib }\citep{matplotlib} \texttt{numpyro}.
Similarly, this work utilized the following julia packages: \texttt{QuadGK.jl} \citep{quadgk}, \texttt{Distributions.jl} \citep{JSSv098i16,Distributions.jl-2019}, \texttt{DataFrames.jl} \citep{JSSv107i04} and \texttt{Roots.jl}\citep{Roots.jl}.

\appendix

\begin{table}[tb]
\centering
\begin{tabular}{lccc}
\hline
Model & $\eta(m_1)$ & Tilt & Role \\
\hline
\textbf{Mass Dependent Free Tilt}   & varying & \textsc{free}  & \textbf{Fiducial} (Sec.~\ref{sec:Results}) \\
Mass Independent Free Tilt & constant & \textsc{free} & Appendix~\ref{sec:ResultsAppx} \\
Mass Dependent Fixed Tilt  & varying & \textsc{fixed} & Appendix~\ref{sec:ResultsAppx} \\
Mass Independent Fixed Tilt & constant & \textsc{fixed} & \cite{Hussain2025} comparison \\
\hline
\end{tabular}
\caption{Summary of the four spin population model variants analyzed in this work. All share the same two-component spin-magnitude mixture (Eq.~\ref{eq:twocomponentmodel}) and differ only in the mass dependence of $\eta$ and the spin-tilt parameterization.}
\label{tab:models}
\end{table}

\section{Details on methods and population models}
\label{sec:MethodsAppx}

\subsection{Population model and hyperpriors}
We model the mass and redshift sectors of the \ac{BBH} population with the
fiducial models of \citet{LIGOScientific:2025pvj}: \textsc{BrokenPowerLawPlusTwoPeaks}
for the primary mass and \textsc{PowerLaw} for the redshift.

The spin magnitudes $(\chi_1, \chi_2)$ are drawn from a mixture of two
bivariate truncated normal distributions over $[0,1]\times[0,1]$, labeled
subpopulations $a$ and $b$, which a posteriori correspond to the slowly and
rapidly spinning subpopulations, respectively. We denote the fraction in
subpopulation $a$ as $\eta$.
Our fiducial model, \textsc{MassDependentFreeTilt}, allows $\eta$ to vary
with the primary mass $m_1$ as
\begin{equation}
\label{eq:eta_model}
\eta(m_1) = \eta_s + \frac{\eta_e - \eta_s}{m_e - m_s}
  \left(\min\!\left[\max\!\left(m_1, m_s\right), m_e\right] - m_s\right),
\end{equation}
a piecewise-linear interpolation from $\eta_s$ to $\eta_e$ over the
transition region $[m_s, m_e]$.

The spin-tilt distribution follows the \textsc{GaussianComponentSpins} tilt
sector of \citet{LIGOScientific:2025pvj}: each cosine tilt is drawn from a
mixture of a uniform (isotropic) component and a truncated Gaussian with
isotropic covariance and a shared, free peak location $\mu_t \equiv
\mu_{\cos\theta_1} = \mu_{\cos\theta_2}$.

Our spin population model has two axes of variation: (i) whether the mixing fraction $\eta$ is constant across masses or follows the piecewise-linear form $\eta(m_1)$ in Eq.~\eqref{eq:eta_model}; and (ii) whether the truncated-normal component for the cosine tilts has a fixed peak at $\cos\theta_i=1$ (\textsc{fixed}, following~\citealt{KAGRA:2021duu}) or a free peak (\textsc{free}, following~\citealt{LIGOScientific:2025pvj}).
The mass-independent variant is recovered by setting $\eta_s=\eta_e\equiv\eta$.
These choices yield four model combinations, summarized in Table~\ref{tab:models}.

\begin{figure}[t]
    \centering
\includegraphics[width=0.48\columnwidth]{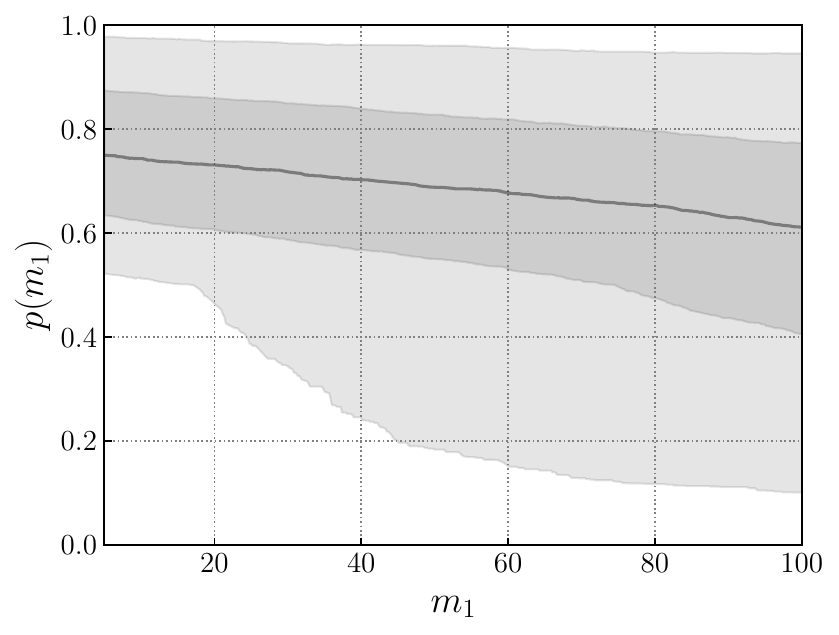}
    \caption{Prior predictive on the mixing fraction $\eta(m_1)$, with the band indicating the 90\% \ac{CI}.}
    \label{fig:prior_predictive}
\end{figure}

Our priors on all hyperparameters are given in Table~\ref{tab:hyperpriors}.
We adopt broad uniform priors on all spin-magnitude parameters, allowing
each subpopulation's mean, width, and inter-spin correlation to vary freely
over their physical ranges. The prior on $\eta_s$ is restricted to $[0.5,1]$
so that subpopulation $a$ is always the low-mass-dominant one by
construction; the characterizations of $a$ as slowly spinning and $b$ as
rapidly spinning reflect our posterior inferences and are not imposed by the
prior. Figure~\ref{fig:prior_predictive} shows the resulting prior
predictive on $\eta(m_1)$.

\begin{table}[tb]
\centering
\begin{tabular}{lll}
\hline
Parameter & Description & Prior \\
\hline
\multicolumn{3}{l}{\it Mixing fraction} \\
$\eta_s$ & fraction at low masses & $U(0.5,1)$ \\
$\eta_e$ & fraction at high masses & $U(0,1)$ \\
$m_s$ & transition start [$M_\odot$] & $U(3,200)$ \\
$m_e$ & transition end [$M_\odot$] & $U(3,200),\; m_s < m_e$ \\
\hline
\multicolumn{3}{l}{\it Subpopulation $a$ (slowly spinning)} \\
$\mu^a_1$, $\mu^a_2$ & spin-magnitude locations & $U(0,1)$ \\
$\sigma^a_1$, $\sigma^a_2$ & spin-magnitude widths & $U(0,1)$ \\
$\rho^a$ & inter-spin correlation & $U(-0.9,0.9)$ \\
\hline
\multicolumn{3}{l}{\it Subpopulation $b$ (rapidly spinning)} \\
$\mu^b_1$, $\mu^b_2$ & spin-magnitude locations & $U(0,1)$ \\
$\sigma^b_1$, $\sigma^b_2$ & spin-magnitude widths & $U(0,1)$ \\
$\rho^b$ & inter-spin correlation & $U(-0.9,0.9)$ \\
\hline
\end{tabular}
\caption{Hyperpriors for the mixing fraction (top) and spin-magnitude subpopulations (bottom).
The mass-independent models use a single constant $\eta \sim U(0.5,1)$ in place of the four mass-dependent parameters.
Spin-tilt hyperparameters share the same uniform priors and are not listed separately.
The labels ``subpopulation $a$'' and ``subpopulation $b$'' are assigned so that $a$ dominates at low masses ($\eta_s \geq 0.5$); the characterizations of $a$ as slowly spinning and $b$ as rapidly spinning reflect our posterior inferences and are not imposed by the model.}
\label{tab:hyperpriors}
\end{table}

\subsection{Summary of TGMM population analysis method}
\label{app:tgmm}

We summarize the \ac{TGMM} population analysis method of \cite{Hussain:2025llf},
highlighting two improvements made for this work.
As part of the standard hierarchical Bayesian inference procedure on \ac{GW}
data~\citep[e.g.][]{Mandel:2018mve,Thrane:2018qnx,Vitale:2020aaz}, we need to
efficiently estimate certain marginalized likelihoods using importance sampling.
The event-level marginalized likelihoods and population-averaged detection
efficiency required for hierarchical inference are both integrals of the form
\begin{equation}
    \label{eq:generic-integral}
    I(\boldsymbol{\Lambda}) = \int p(\boldsymbol{\theta} \mid \boldsymbol{\Lambda})
    \frac{p(\boldsymbol{\theta} \mid \cdot\,)}{W(\boldsymbol{\theta})}\,\mathrm{d}\boldsymbol{\theta}
    \approx \left\langle \frac{p(\boldsymbol{\theta} \mid \boldsymbol{\Lambda})}{W(\boldsymbol{\theta})}
    \right\rangle_{\boldsymbol{\theta} \sim p(\boldsymbol{\theta} \mid \cdot)},
\end{equation}
where $p(\boldsymbol{\theta} \mid \cdot\,)$ is a reference distribution from which
samples are available and $W(\boldsymbol{\theta})$ are their associated weights.
The population posterior is
\begin{equation}
    p(\boldsymbol{\Lambda} \mid \{d_i\}) = \pi(\boldsymbol{\Lambda})\,
    \xi(\boldsymbol{\Lambda})^{-N} \prod_{i}^{N} \mathcal{L}(d_i \mid \boldsymbol{\Lambda}),
    \label{eq:populationposterior}
\end{equation}
where $\pi(\boldsymbol{\Lambda})$ is the hyperprior and $\xi(\boldsymbol{\Lambda})$
is the detection efficiency. The event-level likelihoods and detection efficiency
are
\begin{align}
    \mathcal{L}(d_i \mid \boldsymbol{\Lambda}) &= \int p(\boldsymbol{\theta} \mid
    \boldsymbol{\Lambda})\frac{p(\boldsymbol{\theta} \mid d_i)}{\pi(\boldsymbol{\theta}
    \mid \emptyset)}\,\mathrm{d}\boldsymbol{\theta}
    \approx \left\langle \frac{p(\boldsymbol{\theta} \mid \boldsymbol{\Lambda})}
    {\pi(\boldsymbol{\theta} \mid \emptyset)} \right\rangle_{\boldsymbol{\theta}
    \sim p(\boldsymbol{\theta} \mid d_i)}, \label{eq:eventpoplikelihood} \\
    \xi(\boldsymbol{\Lambda}) &\approx \frac{N_{\rm det}}{N_{\rm draw}}
    \left\langle \frac{p(\boldsymbol{\theta} \mid \boldsymbol{\Lambda})}
    {p(\boldsymbol{\theta} \mid \boldsymbol{\Lambda}_0)} \right\rangle_{\boldsymbol{\theta}
    \sim p_{\rm det}(\boldsymbol{\theta} \mid \boldsymbol{\Lambda}_0)},
    \label{eq:detectionefficiency}
\end{align}
where $\pi(\boldsymbol{\theta} \mid \emptyset)$ is the single-event sampling prior,
$\boldsymbol{\Lambda}_0$ is a fiducial population used for the injection campaign with
$N_{\rm draw}$ simulated signals of which $N_{\rm det}$ are detected, and
$p_{\rm det}(\boldsymbol{\theta} \mid \boldsymbol{\Lambda}_0) \propto
p(\boldsymbol{\theta} \mid \boldsymbol{\Lambda}_0)\,P_{\rm det}(\boldsymbol{\theta})$.
Equations~\eqref{eq:eventpoplikelihood} and \eqref{eq:detectionefficiency} are
both instances of Eq.~\eqref{eq:generic-integral}.

To evaluate Eq.~\eqref{eq:generic-integral} efficiently, we approximate the
reference distribution $p(\boldsymbol{\theta} \mid \cdot\,)$, which represents the event
posteriors or detected injections, as a weighted mixture of truncated multivariate
normal distributions, factored into mass-redshift and spin sectors,
\begin{equation}
    p(\boldsymbol{\theta} \mid \cdot\,) = \sum_k w_k\,
    \mathcal{N}_{[\boldsymbol{a},\boldsymbol{b}]}
    (\boldsymbol{\theta}^{m,z} \mid \boldsymbol{\mu}^{m,z}_k, \boldsymbol{\Sigma}^{m,z}_k)\,
    \mathcal{N}_{[\boldsymbol{a},\boldsymbol{b}]}
    (\boldsymbol{\theta}^\chi \mid \boldsymbol{\mu}^{\chi}_k, \boldsymbol{\Sigma}^{\chi}_k)
    \coloneqq \sum_k w_k\,\phi_k(\boldsymbol{\theta}^{m,z})\,\phi_k(\boldsymbol{\theta}^\chi),
    \label{eq:seperatedtgmmfit}
\end{equation}
where $w_k$ are mixture weights, $\boldsymbol{a}$ and $\boldsymbol{b}$ are the
bounding corners of the truncated domain, and $\phi_k$ is shorthand for each
truncated normal factor. For population models in the spin sector that are themselves mixtures of truncated normals or uniforms (and optionally depend on the values of the mass and redshift $\boldsymbol{\theta}^{m,z}$) the spin overlap integral
\begin{equation}
    \Phi_k(\boldsymbol{\Lambda}^\chi \mid \boldsymbol{\theta}^{m,z}) = \int p(\boldsymbol{\theta}^{\chi} \mid \boldsymbol{\Lambda}^\chi, \boldsymbol{\theta}^{m,z})\,\phi_k(\boldsymbol{\theta}^{\chi})\,
    \mathrm{d}\boldsymbol{\theta}^{\chi}
    \label{eq:spinoverlap}
\end{equation}
can be evaluated semi-analytically, while the mass and redshift sectors of
$I(\boldsymbol{\Lambda})$ are handled with \ac{MC} integration~\citep{Hussain2025,
Hussain:2025llf}. 
The fitting procedure itself follows \cite{Lee2012}, with improvements
from \cite{gmm-anealing,weighted-samples,expectation-conjugate-gradient,
Hussain:2025llf}, and is implemented in
\textsc{truncatedgaussianmixtures}~\citep{truncatedgaussianmixtures}.

In \cite{Hussain:2025llf}, each mass--redshift sample $\boldsymbol{\theta}^{m,z}_i$
was assigned to a single \ac{TGMM} component by drawing a hard one-hot assignment
from the categorical distribution defined by the responsibilities
$\rho_{ik}$, where
\begin{equation}
    \rho_{ik} = \frac{w_k\,\phi_k(\boldsymbol{\theta}^{m,z}_i)}
    {\sum_{k'} w_{k'}\,\phi_{k'}(\boldsymbol{\theta}^{m,z}_i)}.
\end{equation}
Here we instead use the responsibilities directly as soft weights, retaining the
full categorical distribution rather than collapsing it to a single draw.
This is a ``Rao--Blackwellization'' \citep{Rao:1945, Blackwell:1947} of the estimator: replacing a noisy one-hot sample
with its expectation $\rho_{ik}$ reduces the variance of $I(\boldsymbol{\Lambda})$
without introducing bias, and is particularly beneficial for samples whose
responsibilities are spread across multiple components.
The full integral then reduces to an \ac{MC} sum over the mass--redshift sector,
\begin{align}
    \label{eq:finaloptionA}
    I(\boldsymbol{\Lambda}) &= \frac{1}{N_\mathrm{samp}} \sum_i^{N_\mathrm{samp}}
    \sum_k \rho_{ik}\,\frac{p(\boldsymbol{\theta}^{m,z}_i \mid \boldsymbol{\Lambda}^{m,z})}
    {W(\boldsymbol{\theta}^{m,z}_i)\,W(\boldsymbol{\theta}^{\chi}_i)}\,
    \Phi_k(\boldsymbol{\Lambda}^\chi \mid \boldsymbol{\theta}_i^{m,z}).
\end{align}

For our mass dependent fraction, we have a simple form for the semi analytic piece, namely, 
\begin{equation}
    \Phi_k(\boldsymbol{\Lambda}^\chi \mid m_1)
    = \eta(m_1)\,\Phi^a_k + \left[1 - \eta(m_1)\right]\Phi^b_k,
    \label{eq:phifactored}
\end{equation}
where $\Phi^a_k$ and $\Phi^b_k$ are the component-specific integrals that do not depend on $m_1,q$ or $z$.

Finally, we note that the O4a injection campaign~\citep{Essick:2025zed, LIGOScientific:2025yae,
LIGOScientific:2025slb} uses non-uniform spin-sector weights $W(\boldsymbol{\theta}^\chi)$.
We handle this by including the spin-sector weights in the \ac{TGMM} fit, by fitting $p(\boldsymbol{\theta} \mid \cdot\,)$ to samples already weighted by $W(\boldsymbol{\theta}^\chi)$ which absorbs the weight into the \ac{TGMM} coefficients, so that Eq.~\eqref{eq:finaloptionA} remains valid as written.

\section{Further results}
\label{sec:ResultsAppx}

\subsection{Further results for mass-dependent mixing fraction}

\begin{figure}[tb]
    \centering
    \includegraphics[width=\textwidth]{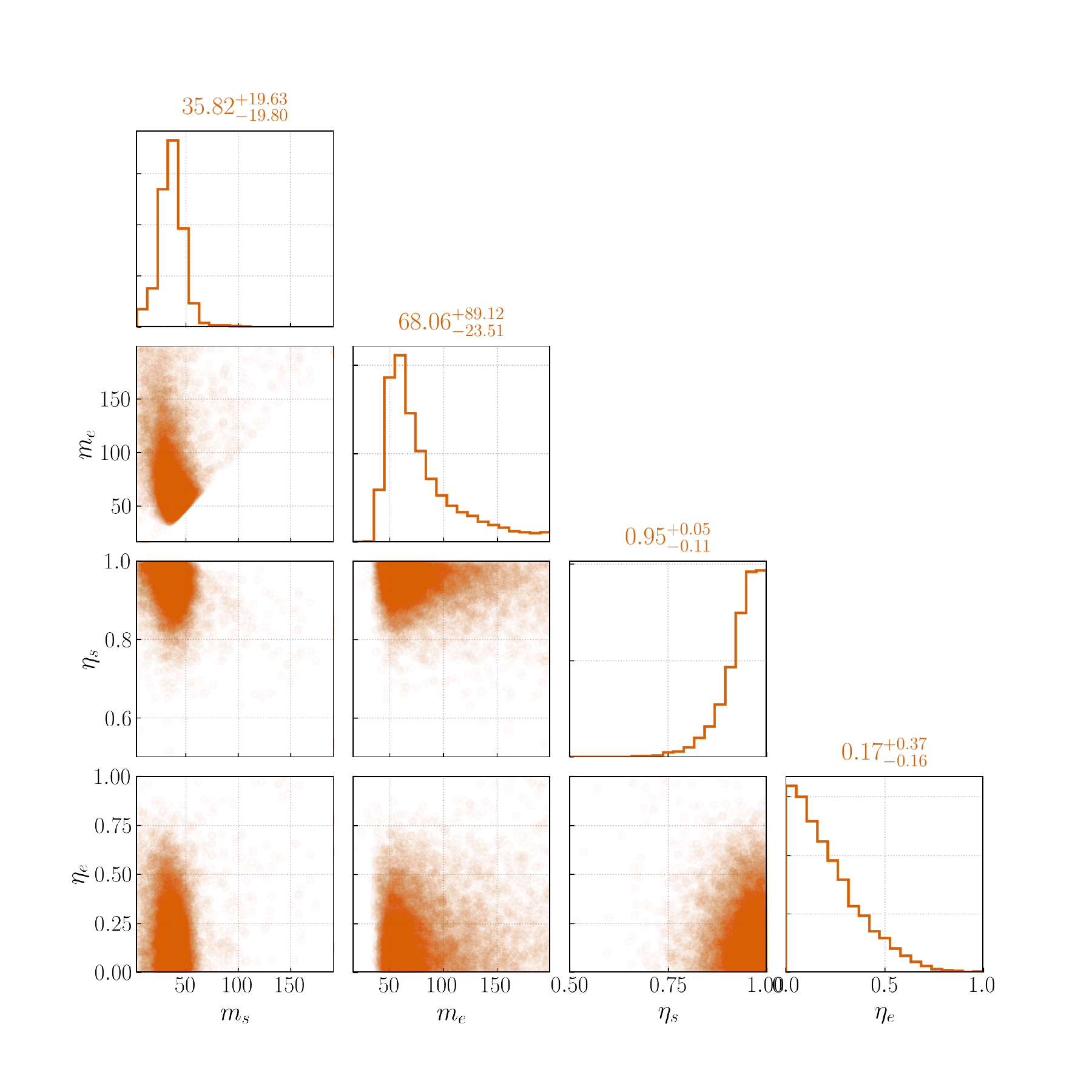}
    \caption{Corner scatter plot illustrating the marginalized hyperposteriors controlling the mixing fraction $\eta(m_1)$, as defined in Equation~\ref{eq:eta_model} and illustrated in Figure~\ref{fig:eta_cartoon}.
    We see negligible posterior support at $\eta_s = \eta_e = 1$, corresponding to a single spin population. Further, we see negligible support along the entire diagonal $\eta_s = \eta_e$, corresponding to a mass-independent mixing fraction.}
    \label{fig:fraction_hyper_param_corner}
\end{figure}

Our fiducial model allows for mass-dependent mixing fraction $\eta(m_1)$ and \textsc{free} tilt model. 
We plot the hyperparameters controlling $\eta(m_1)$ in Figure~\ref{fig:fraction_hyper_param_corner}.

\begin{figure}[tb]
    \centering
    \includegraphics[width=0.5\textwidth]{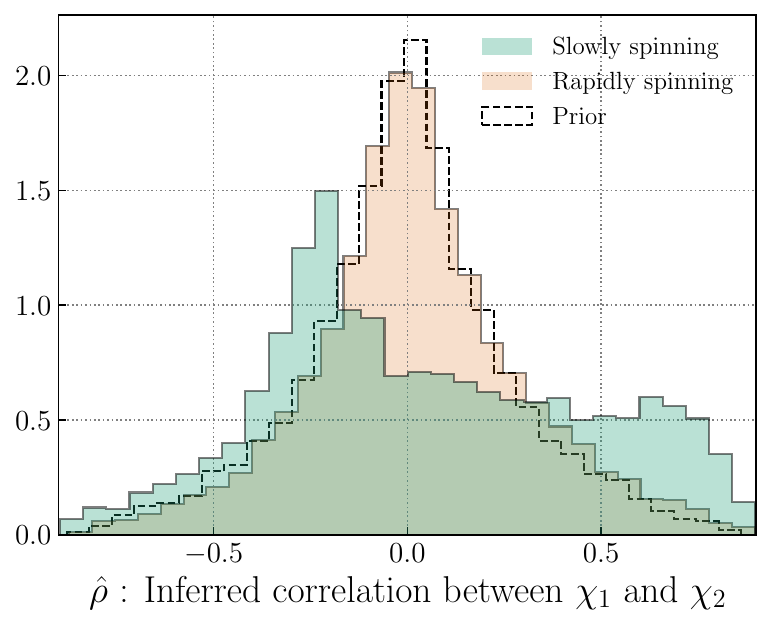}
    \caption{Posteriors of the Pearson correlation coefficients $\hat \rho^a$ and $\hat \rho^b$ for the slowly spinning and rapidly spinning subpopulations, respectively.}
    \label{fig:correlation}
\end{figure}

In addition to the mean and width parameters illustrated in Figure~\ref{fig:fraction_hyper_param_corner}, our spin magnitude model allows for correlation between primary and secondary spins in each subpopulation.
The controlling parameters $\rho^a$ and $\rho^b$ cannot be naively interpreted as the degree of correlation for our truncated normal distributions since their $\boldsymbol{\mu}$ parameters can lie outside the truncation domain, and then the truncation can reshape the distribution. 
For this reason we directly compute the Pearson correlation coefficient $\hat \rho$ for each subpopulation.
This is shown in Figure~\ref{fig:correlation}, along with the induced prior inherited from our priors on the hyperparameters of our model.
We see that the correlation of the rapidly spinning subpopulation is essentially unchanged from the prior.
On the other hand, we see that for the slowly spinning population, the posterior for $\hat \rho^a$ peaks away from the prior, favoring mild anti-correlation between the spins. 
However, it also develops a more extended tail to positive values of $\hat \rho^a$ relative to the prior.
Additional detections are required to better characterize these correlations.

\subsection{Comparison to GWTC-3 results}

\begin{figure}[t]
    \centering
    \includegraphics[width=0.35\columnwidth]{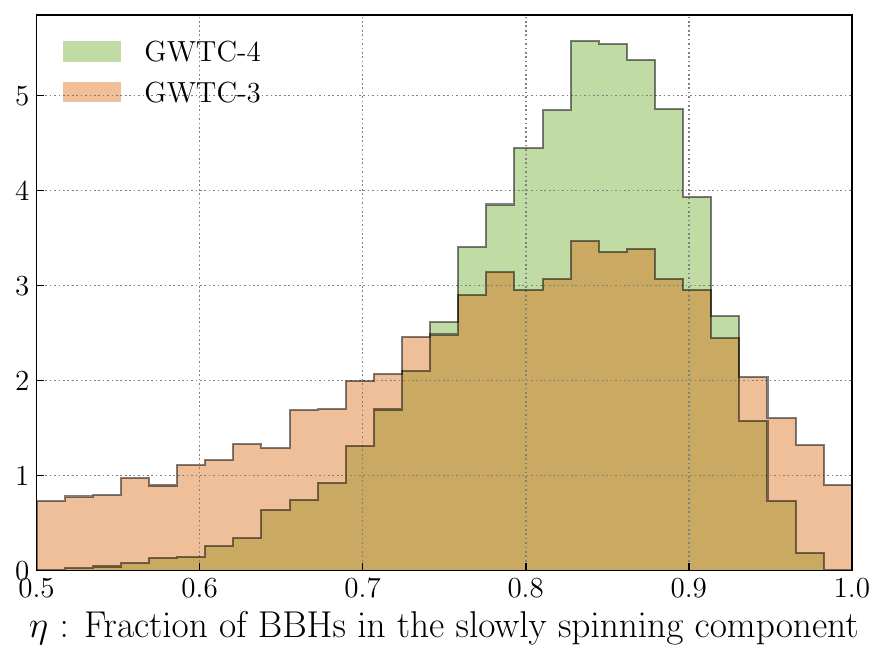}
    \includegraphics[width= 0.275\columnwidth]{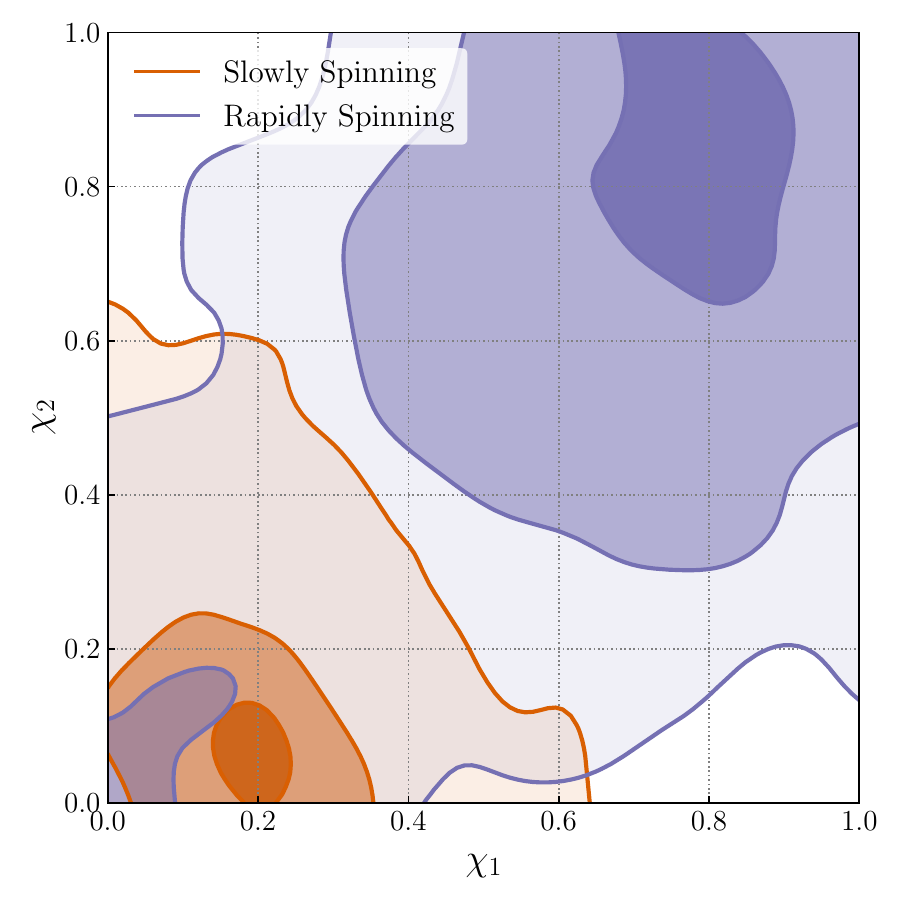}
    \includegraphics[width= 0.275\columnwidth]{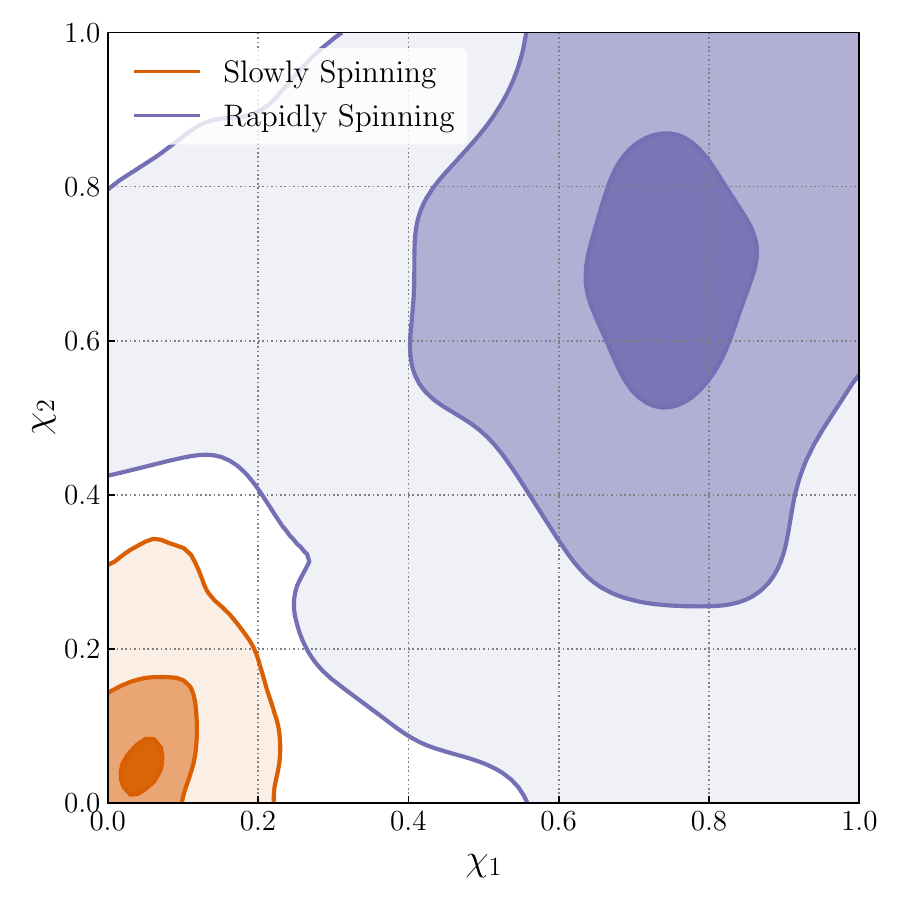}
    \caption{Comparison between our inferences with the GWTC-4 dataset using constant fraction $\eta$ and \textsc{free} tilt model to the results of \cite{Hussain2025} which use the GWTC-3 dataset, \textsc{fixed} tilt model, and mass model of ~\cite{KAGRA:2021duu}.
    {\it Left}: Posterior comparison for $\eta$.
    With GWTC-4 we find strong evidence for two populations. {\it Center}: GWTC-3 result from \cite{Hussain2025} displaying the \ac{PPD} over spin magnitudes for each component. {\it Right}: Similar \ac{PPD} over spin magnitudes using the \textsc{FreeTilt} model with a constant fraction $\eta$ on the GWTC-4 catalog. 
    }
    \label{fig:free_tilt_fixed_fraction_comparison}
\end{figure}

Relative to our previous analysis of~\cite{Hussain2025}, the results presented here utilize more events, a modified mass model, and the \textsc{free} rather than \textsc{fixed} truncated normal component for the cosine tilts. 
To make contact with the previous results following GWTC-3, we fix $\eta$ to be constant across masses but otherwise leave our fiducial model unchanged.
In~\cite{Hussain2025} we preferred two spin populations to one, but the evidence was modest; with constant $\eta$ we find here strong evidence for two populations.
Figure~\ref{fig:free_tilt_fixed_fraction_comparison} shows the posterior over $\eta$ as compared to~\cite{Hussain2025}, with $\eta = \IndependentFreeTiltFraction$ confidently excluding $\eta = 1$. The slowly spinning subpopulation has become increasingly localized near $\chi_i \approx 0$, while the more rapidly spinning subpopulation becomes tighter but is still diffuse. The rapidly spinning subpopulation in GWTC-4 takes up markedly different territory on $\chi_1, \chi_2$ compared to the slowly spinning subpopulation, allowing more confident identification based on spin-magnitudes alone. In comparison, GWTC-3 allowed some label switching behavior due to non-negligible posterior support at $\eta = 0.5$.

\subsection{Fixed tilt results}

\begin{figure}[t]
\includegraphics[width = 0.42\columnwidth]{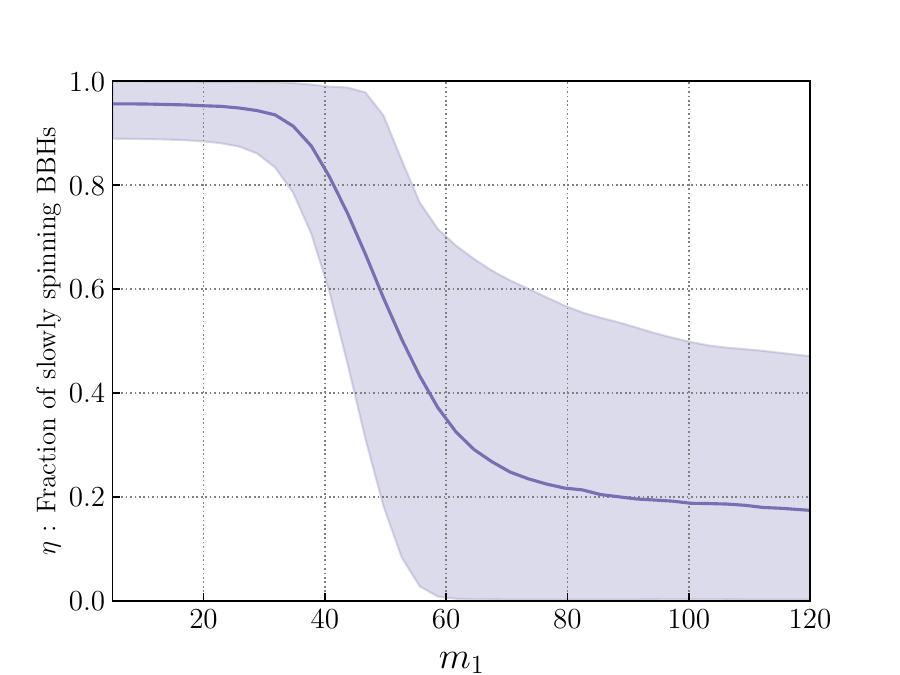}
\includegraphics[width=0.29\columnwidth]{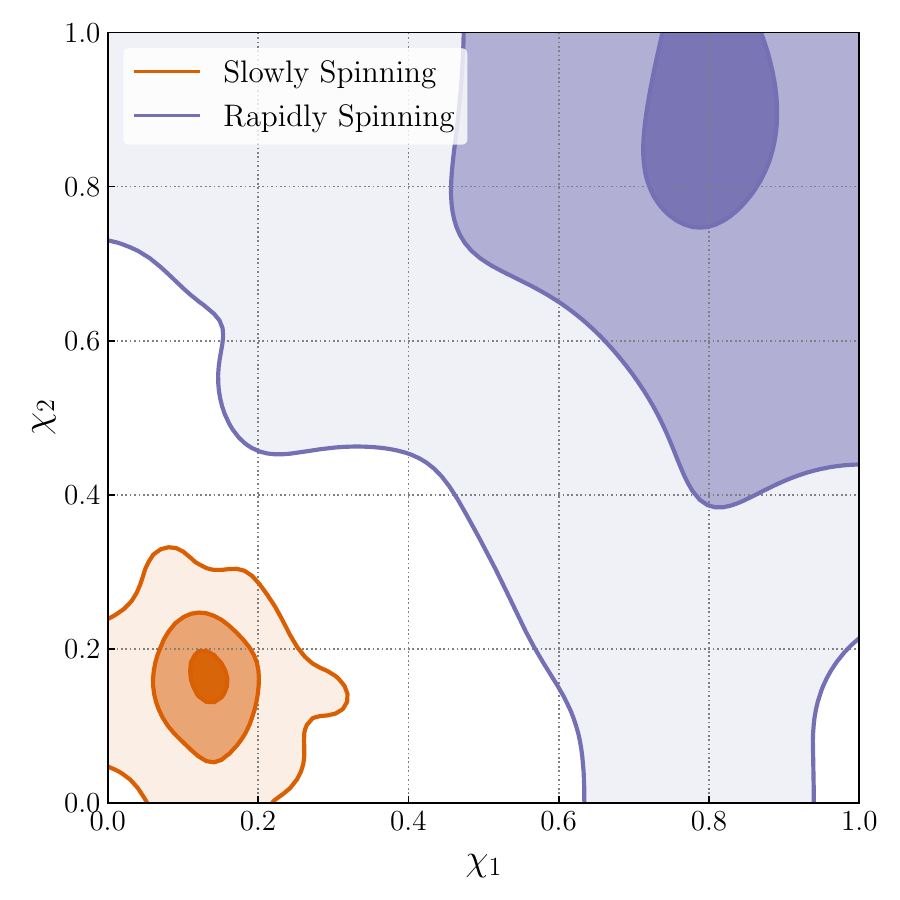}
\includegraphics[width=0.29\columnwidth]{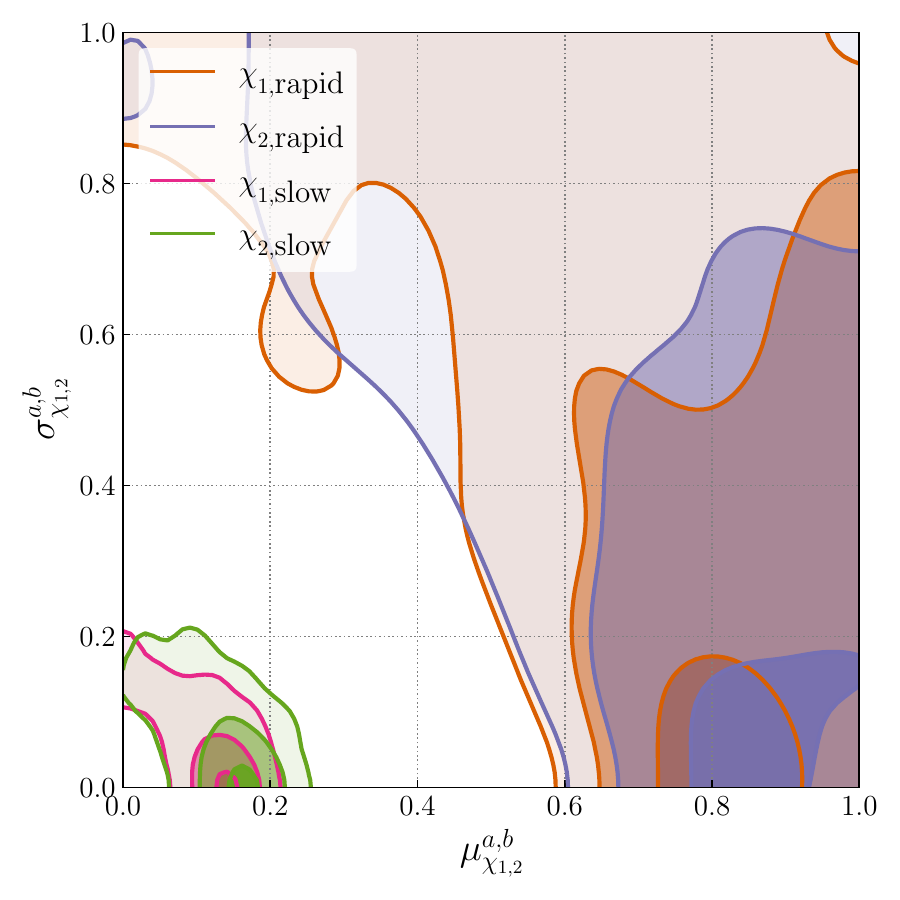}
\caption{Spin magnitude \acp{PPD} for the two subpopulations, individually normalized. In this case we allow $\eta(m_1)$ to vary but use the \textsc{fixed} spin tilt model to peak at $\cos \theta_i = 1$.
    The slowly spinning population is concentrated tightly near $\chi_1 \sim \chi_2 \sim 0.15$, with no evidence for correlation. 
    The rapidly spinning population is concentrated in the region here both black holes are highly spinning, $\chi_1 \sim 0.75$ and $\chi_2 \sim 0.75$, but there is a wide tail towards configurations where the secondary or primary has small spin.}
    \label{fig:fixed_tilt_mass_dependent_results}
\end{figure}

For our fiducial model we adopt the cosine tilt population model of~\cite{LIGOScientific:2025pvj}, which we dub the \textsc{free} tilt model.
In our previous work we adopted the \textsc{fixed} cosine tilt model of~\cite{KAGRA:2021duu}, where the normal component is fixed to peak at $\cos \theta_i = 1$.
Our main results, namely the existence of dominant slowly spinning \ac{BBH} population and a subdominant, rapidly spinning \ac{BBH} population which likely dominates at high masses, remain the same for both tilt models.
However, we find that our inferences about the properties of the slowly spinning population are influenced by our choice of tilt model.

Figure~\ref{fig:fixed_tilt_mass_dependent_results} shows the result of using the \textsc{fixed} tilt distribution.
The main difference relative to our fiducial model is that the slowly spinning distribution moves away from negligible spins, peaking near $\chi_1 \approx \chi_2 \approx 0.15$.
From the hyperposteriors over the spin sector, we see that populations with negligible spins, i.e.~$\mu^a_1 = \mu^a_2 = \sigma^a_1 = \sigma^b_2 = 0$, are disfavored.

In our fiducial \textsc{free} tilt model we find that the truncated normal component peaks near equatorial spin configurations.
Thus the results of the two models appear to be in mild tension; our preferred model serves to minimize the effects of spins aligned with the orbital angular momentum by preferring smaller spins on average and placing more spins in the equatorial plane.
Meanwhile the \text{free} model results hint at a slowly spinning subpopulation with \acp{BBH} with spins $\sim 0.15$ sometimes drawn relatively aligned with the orbit.
However the resulting \acp{PPD} of the spin components of the two models are not radically different: both find that the spin components aligned with the orbital angular momentum, $\chi^a_{i,z}$ peak at zero, are symmetric and narrow, with the \textsc{fixed} model twice as wide as the \textsc{free}. 
For the \textsc{free} model we find $\chi^a_{1,z} = 0.00^{+0.08}_{-0.07}$ and $\chi^a_{2,z} = 0.00^{+0.11}_{-0.10}$, while for the \textsc{fixed} model we find $\chi^a_{1,z} = 0.01^{+0.16}_{-0.16}$ and $\chi^a_{2,z} = 0.02^{+0.18}_{-0.18}$
The distribution of spin components perpendicular to the orbital angular momentum are similar between the model and less well constrained, but systematically slightly wider for the \textsc{free} model.
Our fiducial \textsc{free} model in principle contains our \textsc{free} configuration as a subset of its hyperparameter space, and so we conclude that the populations inferred with the \textsc{fixed} model are not favored by the data.
The differences in the two results hints at more complexity in the slowly spinning population than our model accommodates.

\subsection{Leave one out results}
\label{sec:LeaveOneOutAppdx}

\begin{figure}[tb]
    \centering
    \includegraphics[width=0.54\columnwidth]{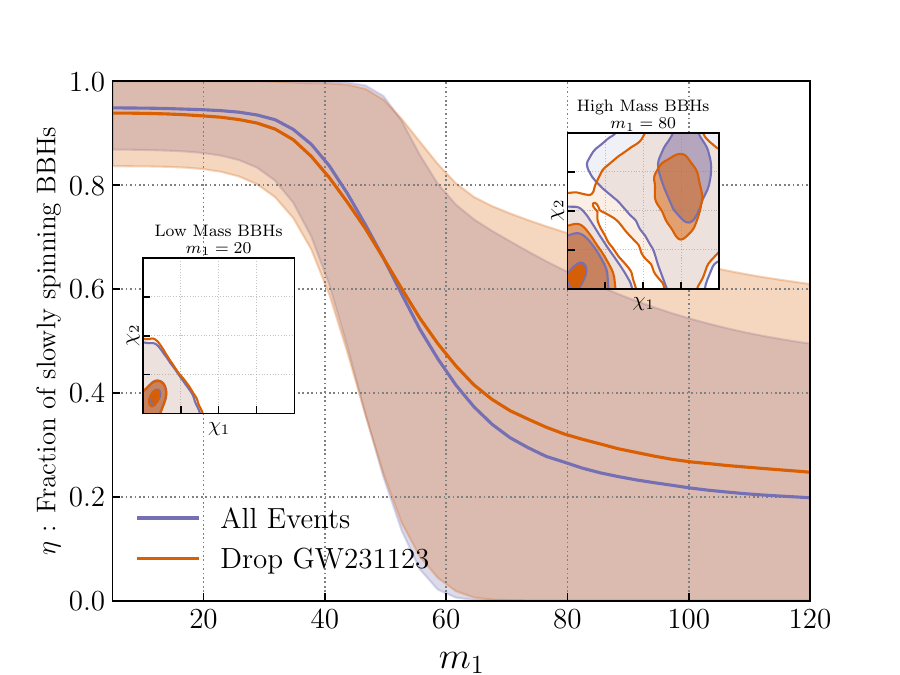}
    \includegraphics[width=0.37\columnwidth]{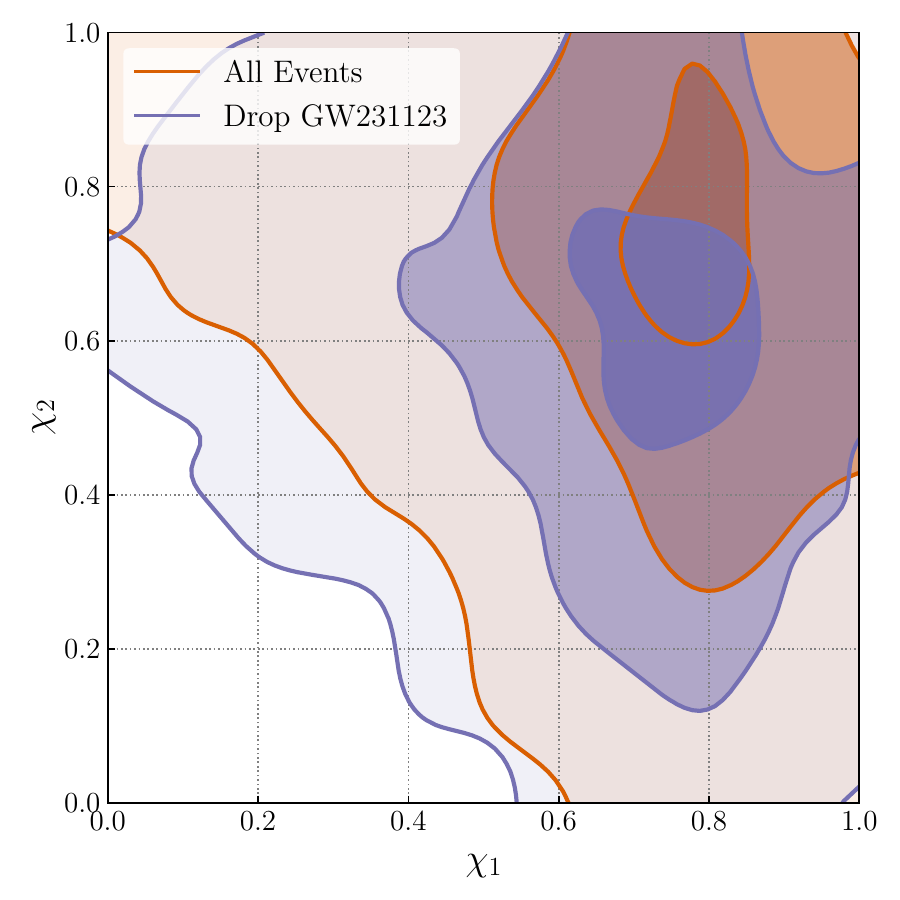}
    \\
    \includegraphics[width=0.54\columnwidth]{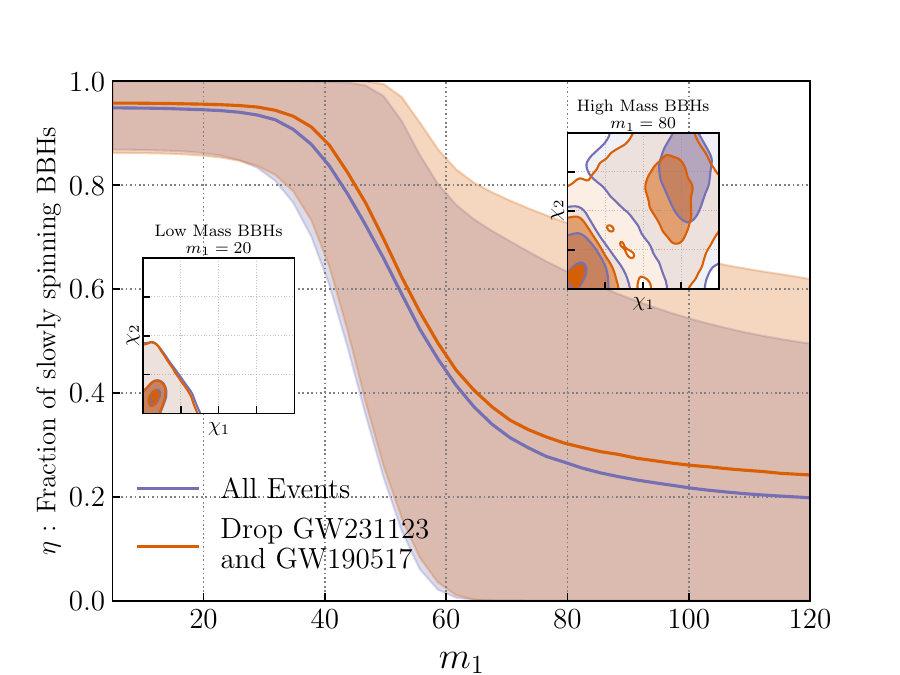}
    \includegraphics[width=0.37\columnwidth]{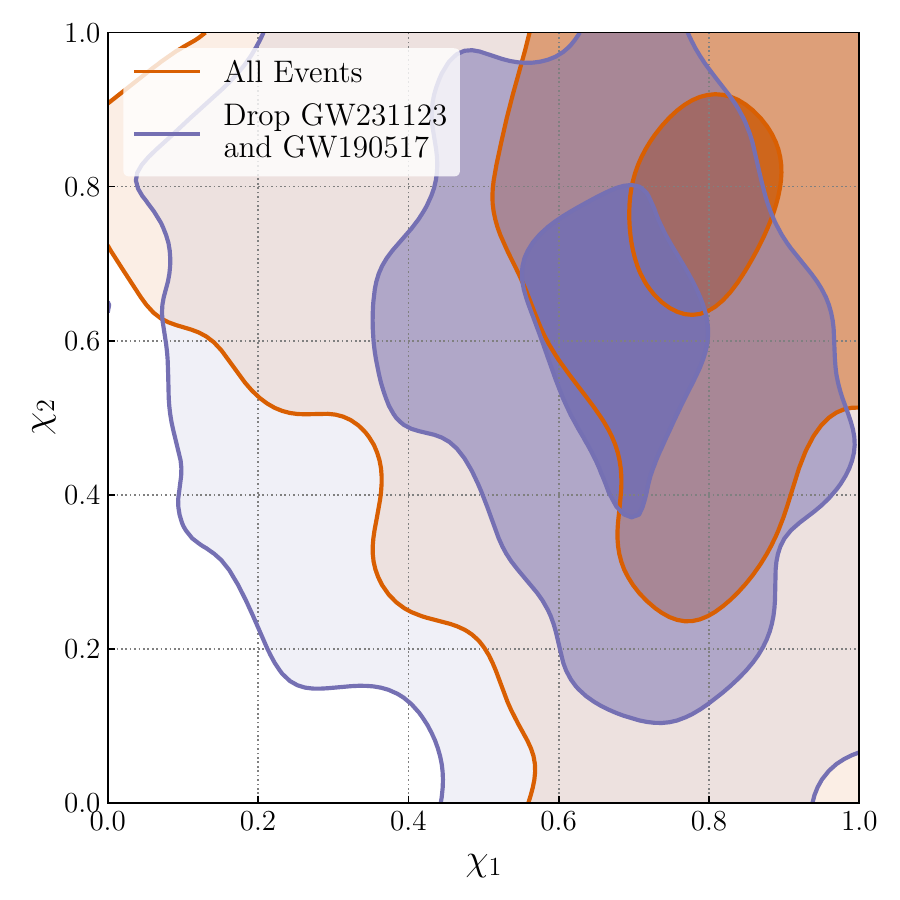}
    \caption{Comparing \acp{PPD} with the inclusion (purple) and exclusion (orange) of GW231123, a high mass and high spin event. \textit{Top left}: \ac{PPD} of the mixing fraction $\eta(m_1)$ with insets showing the inferred spin magnitude distribution at $m_1 = 20\, M_\odot$ and at $80\,M_\odot$, below and above the transition, similar to Fig~\ref{fig:mass_dependent_fraction_ppd}. \textit{Top right}: The \ac{PPD} of the rapidly spinning subpopulation can be seen shifting to lower $\chi_2$ with the removal of GW231123, but still peaks at $\chi_2 > 0$, exhibiting similar tail behaviour towards $\chi_2 = 0$.
    {\it Bottom left}: Same as top left panel, but leaving out both GW231123 and the highly spinning \ac{BBH} GW190517.
    {\it Bottom right}: Sames as top right panel, but leaving out both GW231123 and GW190517.
    }
    \label{fig:leave_one_out}
\end{figure}

Our previous analysis following GWTC-3 was sensitive to excluding the most extremely spinning \ac{BBH} of that dataset, GW190517~\citep{Hussain2025}.
In the GWTC-4 dataset, GW231123, a signal from a massive and highly spinning \ac{BBH}, although how extreme and tightly measured its spin magnitudes are depends on the model applied in inference~\citep{LIGOScientific:2025rsn}.
Here we perform inference with our fiducial model while first excluding GW231123.
Figure~\ref{fig:leave_one_out} compares this leave-one-out analysis to our full result.
While the leave-one-out \ac{PPD} for $\eta(m_1)$ is visibly different than our full analysis, we see that our main conclusions are unchanged: the slowly-spinning population strongly dominates at low masses, with a definite turnover starting at masses $\sim 35 \, M_\odot$. 
At higher masses there is a wider range of mixing fractions, with the rapidly spinning population preferred but with a substantial portion of \acp{BBH} from the slowly spinning population.
The spin distributions themselves are inferred to be very similar with GW231123 left out, although the rapidly spinning component broadens and has somewhat more support for smaller values of $\chi_2$.
The same is conclusions hold if we leave out both GW231123 and GW190517, as seen in Figure~\ref{fig:leave_one_out}.

\bibliographystyle{aasjournal}
\bibliography{refs.bib}

\end{document}